\documentclass[aps,preprint,amsmath,amssymb,showpacs,prd]{revtex4-1}
\usepackage{graphicx}
\usepackage{epstopdf}
\usepackage{amsmath}
\usepackage{amsfonts}
\usepackage{amssymb}

\begin{document}
\title
{\Large {\bf Searching for new light gauge bosons at $e^+e^-$ colliders}}

\author{ I. Alikhanov}
\email[]{E-mail: ialspbu@gmail.com}

\affiliation{Institute for Nuclear Research of the Russian Academy of Sciences, Moscow 117312, Russia\\
       Institute of Applied Mathematics and Automation, Nalchik 360000, Russia}

\author{E. A. Paschos}
\email[]{E-mail: paschos.e@gmail.com}

\affiliation{Department of Physics, TU-Dortmund, 44221 Germany}


\begin{abstract}
Neutral gauge bosons beyond the Standard Model are becoming interesting as possible mediators to explain several experimental anomalies. They have small masses, below one GeV,  and are referred to as dark photons, $U$, $A'$ or $Z'$ bosons. Electron--positron collision experiments at the B-factories provide the most straightforward way to probe bosons of this kind. In the present article we study production of the bosons at $e^+e^-$ colliders operating at GeV center-of-mass energies. We have studied two channels: $e^+e^-\rightarrow \gamma Z'$ and $e^+e^-\rightarrow e^+e^-Z'$. Analytic expressions for the cross sections and various observables such as the energy spectra of the produced bosons and the final electrons from the $Z'$ decays are derived. We have also studied the transverse momentum distribution of the bosons and the spatial  distribution of the $Z'\rightarrow e^+e^-$ decay vertices. It is shown that these distributions provide distinct signatures of the bosons in $e^+e^-\rightarrow\gamma Z'$. The reaction $e^+e^-\rightarrow e^+e^-Z'$ becomes important at small $Z'$ scattering angles where its contribution to the overall yield may be larger by orders of magnitude compared to $e^+e^-\rightarrow\gamma Z'$. The standard processes $e^+e^-\rightarrow\gamma\gamma$ and $e^+e^-\rightarrow e^+e^-\gamma$ that  lead to the same signal are considered.
We include numerical predictions for the production rates at the energy $\sqrt{s}=10.5$~GeV. The case with a light scalar boson is also discussed. The calculations are performed in detail and can be useful for additional studies.

\end{abstract}
\maketitle %

%%%%%%%%%%%%%%%%%
\section{Introduction}
%%%%%%%%%%%%%%%%
The notion of gauge bosons has become an integral part of particle physics. A gauge boson is an electrically neutral or charged particle with spin  one responsible for transmission of forces in a theory. Well known representatives with precisely established properties are the photon, $Z^0$ and $W^\pm$.

Many extensions of the Standard Model accommodate new gauge bosons. After the electroweak $SU(2)\times U(1)_Y$ model was proposed, there appeared numerous alternative theories with additional $U(1)'$ symmetries leading to associated new neutral bosons~\cite{Hewett:1988xc,Langacker:1980js,Langacker:2008yv}. 
Production of heavy mass states in the TeV region have been studied~\cite{Altarelli,Aliev} and searched for directly at the LHC in the ATLAS and CMS experiments which put stringent limits on their masses and couplings to particles of the Standard Model~\cite{Aad:2015osa,CMS:2016zxk,ATLAS:2016cyf,CMS:2016abv}. These bosons have also been indirectly probed using high-precision electroweak data~\cite{Erler:2009jh}.

Apart from the heavy mediators, models with much lighter gauge bosons of masses around one GeV or even a few tens of MeV are extensively discussed in articles and are popular today~\cite{Alexander:2016aln}. They are often referred to as dark photons, $U$, $A'$ or $Z'$ bosons. Such models have been widely studied for various reasons. In particular, their existence was assumed in order to reconcile the measured value of the anomalous magnetic moment of the muon with theoretical calculations~\cite{Gninenko:2001hx,Baek:2001kca,Ma:2001md,Pospelov:2008zw,Heeck:2011wj,Krasnikov:2017dmg}. The bosons were introduced to account for some cosmological and astrophysical observations~\cite{Fayet:2007ua, ArkaniHamed:2008qn,Petraki:2014uza,Foot:2014uba,Araki:2015mya,Baek:2015fea,Ko:2016uft}. 
Possible impact of a new gauge interaction with a light mediator on rare kaon decays~\cite{Pospelov:2008zw,Davoudiasl:2014kua,Chen:2016kxw,Campos:2017dgc} and the Higgs boson decay~\cite{Aad:2015sva,Davoudiasl:2012ag,Davoudiasl:2013aya,Lee:2013fda,Curtin:2014cca,Lee:2016ief} has been investigated.

Recently, the Atomki Collaboration reported the observation of an anomalous bump in the angular and invariant mass distributions of electron--positron pairs emitted in nuclear transitions, $^8\text{Be}^*\rightarrow{}^8\text{Be}+e^+e^-$, with high statistical significance of $6.8\sigma$~\cite{Krasznahorkay:2015iga}. The known nuclear theory predicts no such a bump. A possible explanation of this anomaly can be an additional channel with the emission of a light neutral vector boson, subsequently decaying into a $e^+e^-$ pair~\cite{Krasznahorkay:2015iga,Feng:2016jff,Feng:2016ysn,Krasznahorkay:2017qfd,Krasznahorkay:2017gwn}. To describe the experimental distributions, the new boson should have mass $m_Z=16.70\pm0.35(\text{stat})\pm0.50(\text{sys})$ MeV~\cite{Krasznahorkay:2015iga}. In the last year, there appeared many articles devoted to this hypothetical 17 MeV boson~\cite{Gu:2016ege,Chen:2016dhm,Liang:2016ffe,Jia:2016uxs,Kitahara:2016zyb,Ellwanger:2016wfe,Chen:2016tdz,Seto:2016pks,Kozaczuk:2016nma,Zhang:2017zap,DelleRose:2017xil,Fornal:2017msy}. First limits on its coupling to electrons have been set in the NA64 experiment at the CERN SPS~\cite{Banerjee:2018vgk}.

Searches for a light boson in $\pi^0\rightarrow Z'+\gamma$ require that $Z'$ should couple to $u$ and $d$ quarks very weakly~\cite{Batley:2015lha}, which means that such a boson should be, as usually dubbed in the literature, "protophobic". On the other hand, if the Atomki anomaly is a manifestation of new physics, then the boson could be produced in a reversed process, for example in $e^+e^-\rightarrow \gamma Z'$~\cite{Fayet:2007ua,Boehm:2003hm,Borodatchenkova:2005ct,Chen:2016dhm}. Electron--positron collisions are the most straightforward reactions to probe $Z'$s~\cite{Essig:2009nc,Essig:2013vha,Lees:2014xha,Lees:2017lec}.  At the same time, one should keep in mind that the value of the coupling must be compatible with other measurements in which $Z'$ may contribute as the electron magnetic dipole moment~\cite{Patrignani:2016xqp}, beam dump experiments and $\nu e$ scattering~\cite{Deniz:2009mu}.

In this paper we focus our attention on the search for the new light gauge bosons at $e^+e^-$ colliders in the reactions $e^+e^-\rightarrow \gamma Z'$ and  $e^+e^-\rightarrow e^+e^-Z'$. 
Introducing a general $V-A$ interaction we present a detailed analytic study of the corresponding production cross sections and emphasize experimental signatures. For completeness, we also investigate a scalar theory considering the production of a spinless light boson in the same channels.

The article is organized as follows. In Section~\ref{sec1d} we study the reaction $e^+e^-\rightarrow \gamma Z'$ and derive distributions of various observables, such as the transverse momentum distribution of the bosons, the distribution of the $Z'\rightarrow e^+e^-$ decay vertex positions, the energy spectrum of the final electrons and positrons from the boson decays. A similar analysis in the framework of the equivalent photon approximation is performed for $e^+e^-\rightarrow e^+e^- Z'$ in Section~\ref{sectionm}. In Section~\ref{sec2} we consider the production of a scalar boson. Section~\ref{concl} summarizes our results. Four Appendices~are devoted to calculations in Sections~\ref{sec1d} and~\ref{sectionm}.

%%%%%%%%%%%%%%%%%%%%%%%%%%%%%%%%%%%%%%%%%
\section{Vector bosons in $e^+e^-\rightarrow \gamma Z'$ reaction\label{sec1d}}
%%%%%%%%%%%%%%%%%%%%%%%%%%%%%%%%%%%%%%%%%

Existing constraints to the parameters of the light $Z'$ bosons with the mass $m_Z\sim 10$~MeV lead to extremely narrow  partial widths for the decay $Z'\rightarrow e^+e^-$, less than $0.1$ eV. This makes their detection at collider experiments as a peak in the  $e^+e^-$ invariant-mass distribution a challenging problem.  It is therefore important to investigate other possibilities for identification of the bosons. For example, a way  to  search
for  narrow  resonances,  and  specifically  $Z'$,  coupled  to $e^+e^-$ pairs has been proposed in~\cite{Nardi:2018cxi}.
In this chapter we study distributions of several observables in $e^+e^-\rightarrow \gamma Z'$.

%%%%%%%%%%%%%
\subsection{Transverse momentum distribution}
%%%%%%%%%%%%%

\begin{figure}
\includegraphics[width=1.\textwidth]{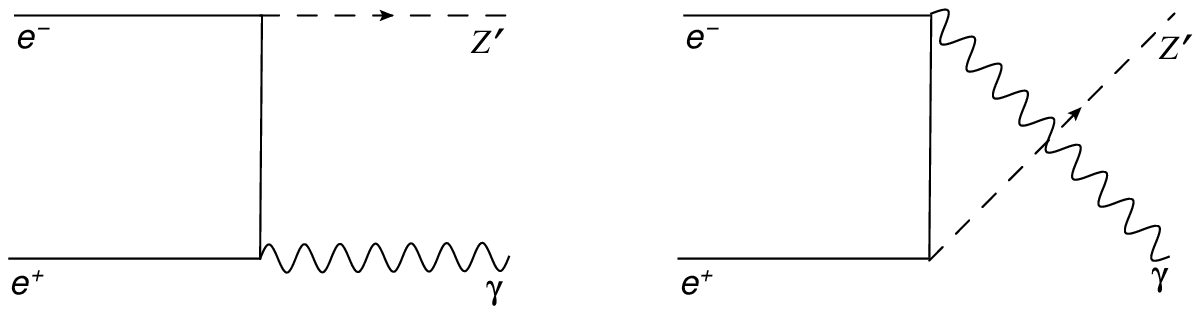}
\caption{Leading Feynman diagrams that contribute to reaction $e^+e^-\rightarrow \gamma Z'$.}
\label{fig1p}
\end{figure}

Consider the production of electrically neutral vector boson $Z'$ in the reaction

\begin{equation}
e^+e^-\rightarrow \gamma Z'.\label{reac22}
\end{equation}
This channel may be favorable for searches at electron--positron colliders~\cite{Fayet:2007ua,Boehm:2003hm,Borodatchenkova:2005ct,Chen:2016dhm}. Assume the interaction

\begin{equation}
\mathcal{L}_e=-e\varepsilon \,\bar u_e\gamma_\mu(g_V-g_A\gamma_5) u_e Z'^{\mu},\label{interact2}
\end{equation}
where $e$ is the elementary electric charge, $\varepsilon$ denotes the coupling strength of the boson to the vector current.
The lowest-order Feynman diagrams contributing to this process (see Fig.~\ref{fig1p}) lead to the following square of the amplitude averaged over the initial and summed over final spins:
%Noting that $e^+e^-\rightarrow \gamma Z'$  is related to reaction $\gamma e^-\rightarrow e^- Z'$ by crossing symmetry we can immediately obtain the corresponding amplitude squared for~\eqref{reac22}. We need to exchange $s\leftrightarrow t$ in~\eqref{amplv} so that

\begin{equation}
|M|^2(s,t,u)=2e^4 \varepsilon^2 (g_V^2+g_A^2) \left(\frac{u}{t}+\frac{t}{u}+\frac{2 m_Z^2s}{ut}\right).\label{amplvl}
\end{equation}
Throughout the article we adopt the Mandelstam variables $s$, $t$ and $u$. Note that $\sqrt{s}$ is the center-of-mass (cms) energy. Also note that the electron mass is neglected because the standard experimental conditions imply $s\gg m_e^2$.

\begin{figure}
\includegraphics[width=1.\textwidth]{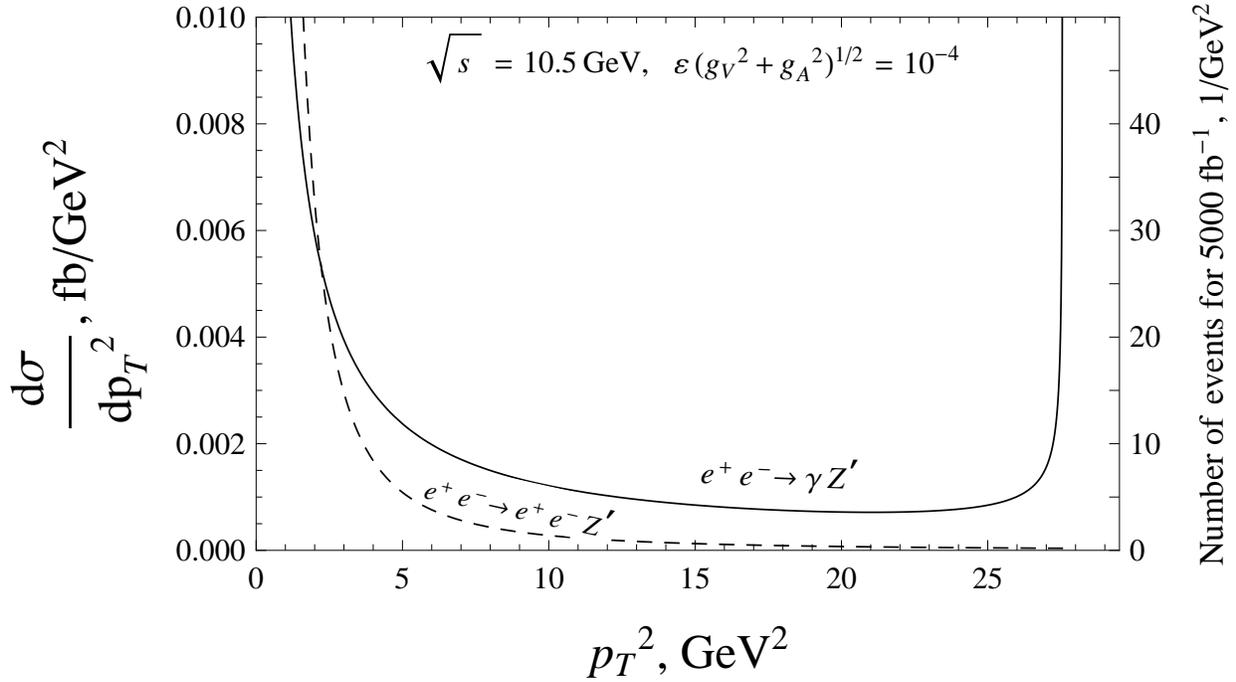}
\caption{Transverse momentum distributions of the $Z'$ bosons produced in $e^+e^-\rightarrow \gamma Z'$ (solid curve) and in $e^+e^-\rightarrow e^+e^- Z'$ (dashed curve) at the cms energy $\sqrt{s}=10.5$ GeV. The coupling is fixed to $\varepsilon(g_V^2+g_A^2)^{1/2}=10^{-4}$. The right-hand-side of the figure is labeled with the number of events corresponding to an integrated
luminosity of 5000 fb$^{-1}$. Note that the distributions are independent on the bosons mass as long as $m_Z\ll\sqrt{s}$.}
\label{fig2p}
\end{figure}

It is interesting to investigate the behavior of the transverse momentum distribution of the produced bosons. 
The cross section differential in $p_T^2$, which actually represents the distribution, generally reads  

\begin{equation}
\frac{d\sigma_{\gamma Z}}{dp_T^2}=\frac{1}{16\pi s^2\sqrt{1-\dfrac{4 p_T^2}{s}}}|M|^2(s,p_T^2),\label{subptf}
\end{equation}
where $|M|^2$ is now the function of two independent variables, $s$ and $p_T^2$.
Using that 

\begin{equation}
p_T^2=\frac{ut}{s}\label{ptuts}
\end{equation}
one can rewrite~\eqref{amplvl} as 

\begin{equation}
|M|^2(s,p_T^2)=2e^4 \varepsilon^2 (g_V^2+g_A^2)  \frac{s}{p_T^2}\left(1-\frac{2p_T^2}{s}+\frac{m_Z^4}{s^2}\right).\label{amplvl3}
\end{equation}
Since we are interested in bosons with masses of the order of several tens of MeV or lighter, the term $m_Z^4/s^2$ in~\eqref{amplvl3} can also be safely omitted. Then, from~\eqref{subptf} and~\eqref{amplvl3}  we obtain

%\begin{equation}
%\frac{d\sigma_{\gamma Z}}{dp_T^2}=\frac{2\pi\alpha^2\varepsilon^2(g_V^2+g_A^2)}{sp_T^2\sqrt{1-\dfrac{p_T^2}{p^2}}}\left(1-\frac{p_T^2}{2p^2}\right),\label{subptf2}
%\end{equation}

\begin{equation}
\frac{d\sigma_{\gamma Z}}{dp_T^2}=2\pi\alpha^2\varepsilon^2(g_V^2+g_A^2)\frac{1}{sp_T^2\sqrt{1-\dfrac{p_T^2}{p^2}}}\left(1-\frac{p_T^2}{2p^2}\right),\label{subptf2}
\end{equation}
where $\alpha=e^2/4\pi$ is the fine structure constant, $p=\sqrt{s}/2$ is the magnitude of the $Z'$ boson three-momentum (provided $m_Z^2\ll s$).
A plot of the transverse momentum distribution~\eqref{subptf2} for $\sqrt{s}=10.5$ GeV is shown in Fig.~\ref{fig2p}. Note that the distribution grows up rapidly  at $p_T^2\rightarrow p^2$. In this case nearly all the momentum of the produced boson is carried away in the transverse direction. This feature should lead to an enhancement of $Z'$ events in the transverse plane which goes through the collision point of the $e^+$ and $e^-$ beams. For $\varepsilon(g_V^2+g_A^2)^{1/2}=10^{-4}$ there may appear about $20$ bosons in this region for an integrated luminosity of $5000$~fb$^{-1}$. In particular, 514 fb$^{-1}$ of data collected by the BaBar experiment~\cite{Lees:2014xha} can be sensitive to such  bosons  with the mass $\lesssim17$ MeV and coupling $\sim10^{-3}$.

If needed, the formalism of this paper allows one to easily incorporate interactions of $Z'$ with other particles. For example, if the bosons leave the experimental setup undetected due to invisible decays~\cite{Lees:2017lec,Araki:2017wyg,Chen:2017cic,Gninenko:2017yus}, say, due to open $Z'\rightarrow\nu\bar\nu$ channels, there should be observed an enhancement of the $e^+e^-\rightarrow \gamma~+~missing~p_T$ events in this plane and in its vicinity.  Since the production of the bosons and their subsequent  $Z'\rightarrow\nu\bar\nu$ decays are independent processes, the missing $p_T^2$ distribution will read

\begin{equation}
\frac{d\sigma_{\gamma Z}}{dp^2_{Tmiss}}=\text{Br}(Z'\rightarrow\nu\bar\nu)\left.\frac{d\sigma_{\gamma Z}}{dp^2_{T}}\right\vert_{p_T^2=p^2_{Tmiss}},\label{dee_brp}
\end{equation}   
where $\text{Br}(Z'\rightarrow\nu\bar\nu)$ is the branching ratio for $Z'\rightarrow\nu\bar\nu$. Also note that~\eqref{subptf2} and~\eqref{dee_brp} are invariant under Lorentz transformations along the beams.

%%%%%%%%%%%%%%%%%%%%%%%%%%
\subsection{Spatial distribution of the $Z'\rightarrow e^+e^-$ decay vertices}
\begin{figure}
\includegraphics[width=1.\textwidth]{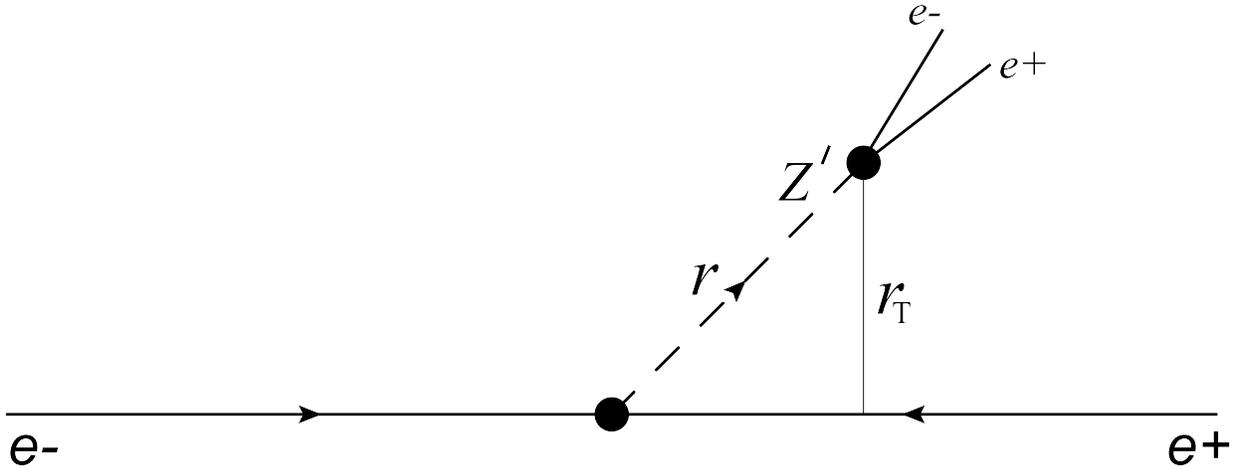}
\caption{Distance $r_T$ traveled by a $Z'$ boson from its production to the decay in the direction transverse to the colliding $e^-$ and $e^+$ beams. The $z$ axis points along the electron beam.}
\label{fig3p}
\end{figure}

%%%%%%%%%%%%%%%%%%%%%%%%
The produced bosons decaying in the detector may leave $e^+e^-$ vertices or single-track electromagnetic showers whose positions can be measured. We study next how these vertices are distributed with respect to the incident electron beam. For this purpose we calculate the distribution of the distances between the beam axis and the $Z'\rightarrow e^+e^-$ decay vertices, as shown in Fig.~\ref{fig3p}.  In other words we wish to find the density of the vertices in the transverse direction.
The distance $r$ traveled by a $Z'$ boson from the interaction point of the beams to the decay point can be defined as

\begin{equation}
r=v\gamma(v) \tau,\label{dist}
\end{equation}
where $v$, $\gamma(v)$ and $\tau$ are the velocity, the Lorentz factor and the mean lifetime at rest of $Z'$, respectively. We need the transverse component of the three-vector ${\bf r}=\textit{\textbf{v}}\gamma(v)\tau$.
We decompose $\textit{\textbf{v}}\gamma(v)$ into transverse and longitudinal components~\cite{byckling}

\begin{equation}
\textit{\textbf{v}}\gamma(v) =\left(\frac{p_T}{m_Z},\,\, \frac{p_L}{m_Z}\right)\label{vel}
\end{equation}
and obtain the distance traveled by the boson in the transverse direction to be

\begin{equation}
r_T=\frac{p_T}{m_Z\Gamma},\label{dist2}
\end{equation}
where $\Gamma=1/\tau$ is the partial width for the $Z'\rightarrow e^+e^-$ decay (see Appendix~\ref{app:decay})

\begin{equation}
\Gamma=\frac{\alpha\varepsilon^2(g_V^2+g_A^2)}{3}m_Z. \label{decayg5h}
\end{equation}
The transverse momentum distribution is related to the distribution in space as follows

\begin{equation}
\frac{d\sigma_{\gamma Z}}{dr_T}=\frac{d\sigma_{\gamma Z}}{dp_T^2}\left|\frac{d p_T^2}{dr_T}\right|
\end{equation}
so that using~\eqref{subptf2} and~\eqref{dist2} gives

\begin{figure}
\includegraphics[width=1.\textwidth]{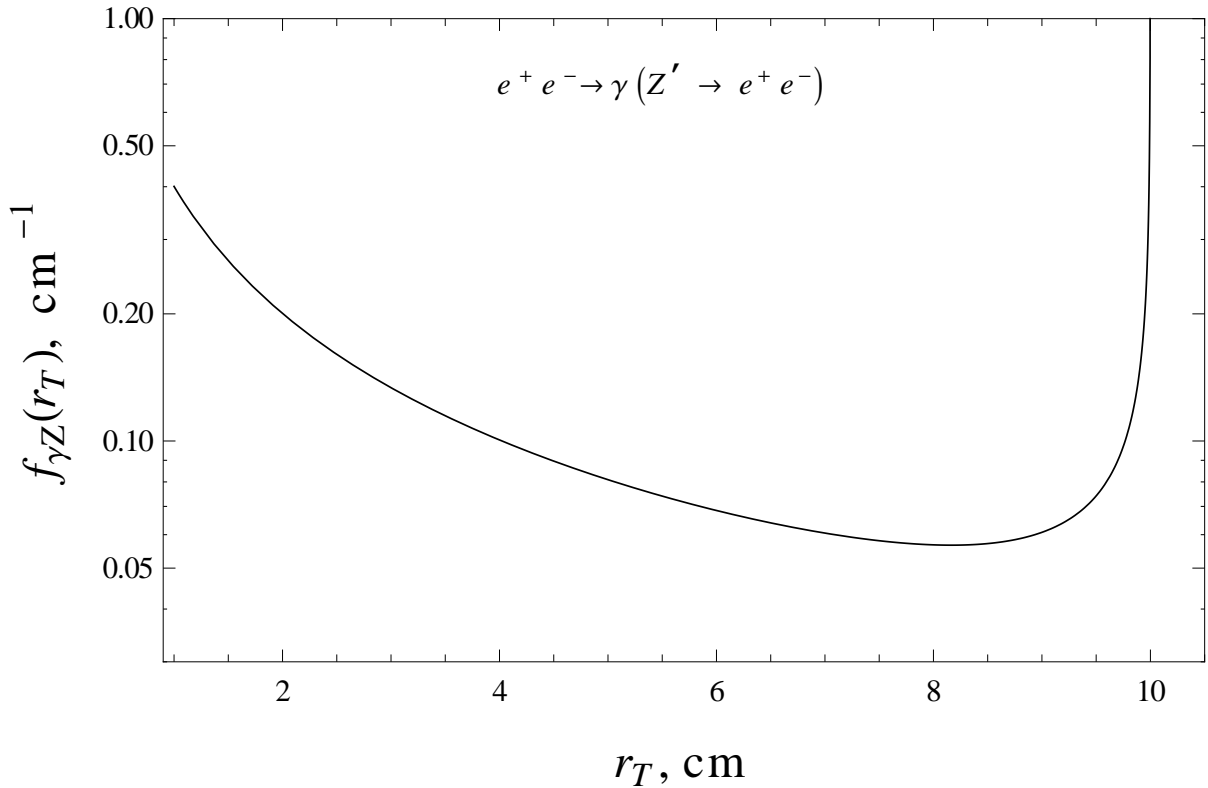}
\caption{The probability density that a $Z'\rightarrow e^+e^-$ decay vertex will appear at the distance between $r_T$ and $r_T+dr_T$ from beam axis for $r_m=1$ cm and $r=10$ cm.  The bosons are produced in $e^+e^-\rightarrow\gamma Z'$.}
\label{fig4p}
\end{figure}

\begin{figure}
\includegraphics[width=1.\textwidth]{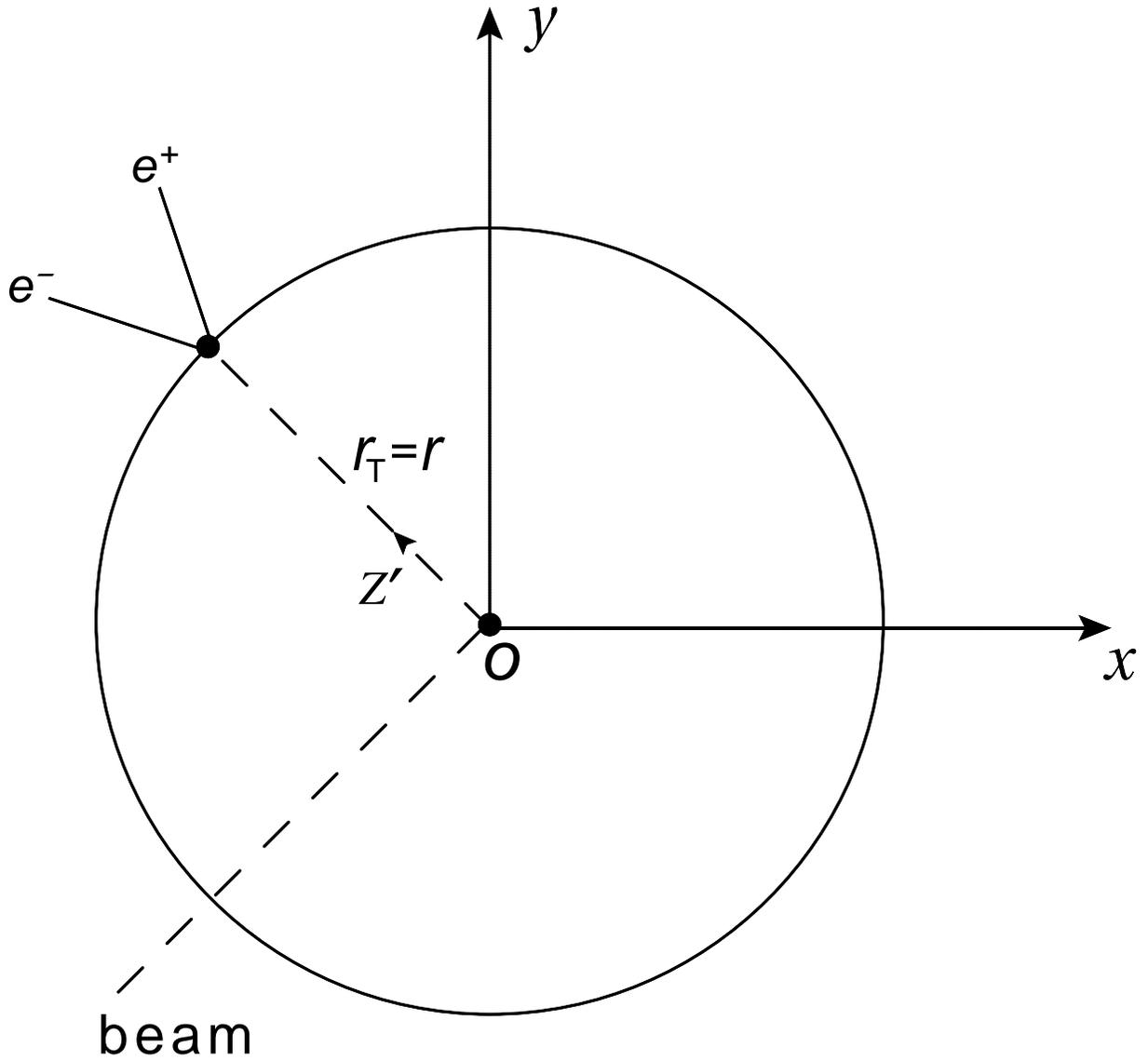}
\caption{A schematic illustration of the ring formed by the  $Z'\rightarrow e^+e^-$ decay events at distance $r_T=r$ in the transverse plane ($x$, $y$).  The bosons are produced in $e^+e^-\rightarrow\gamma Z'$. The electron beam is normal to the plot and directed along the $z$ axis (dashed line). The origin $O$ coincides with the interaction point of the colliding $e^+$ and $e^-$ beams.}
\label{fig5p}
\end{figure}

\begin{figure}
\includegraphics[width=1.\textwidth]{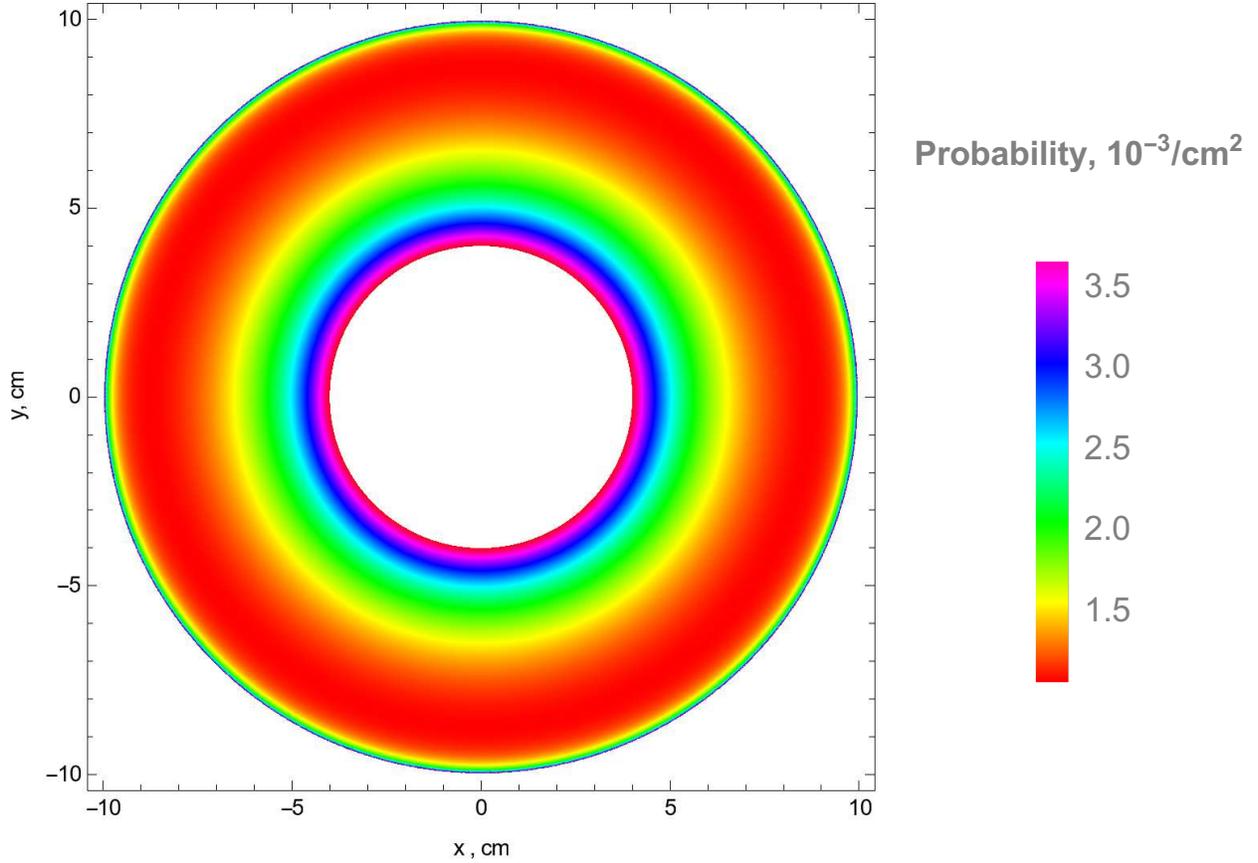}
\caption{Distribution of the probability to observe a $Z'\rightarrow e^+e^-$ decay vertex in the transverse plane for $r_m=1$ cm, $r=10$ cm. The bosons are produced in $e^+e^-\rightarrow\gamma Z'$. The origin coincides with the interaction point of the colliding $e^+$ and $e^-$ beams. The region $\sqrt{x^2+y^2}\leq 4$ cm is cut out.}
\label{fig6p}
\end{figure}

%\begin{equation}
%\left|\frac{d p_T^2}{dr_T}\right|=2r_Tm_Z^2\Gamma^2,\label{dist3}
%\end{equation}
%so that from~\eqref{subptf2} and~\eqref{dist3} we have

%\begin{equation}
%\frac{d\sigma_{\gamma Z}}{dr_T}=\frac{4\pi\alpha^2\varepsilon^2(g_V^2+g_A^2)}{s\,r_T\sqrt{1-\dfrac{r_T^2}{r^2}}}\left(1-\dfrac{r_T^2}{2r^2}\right),\label{dist4}
%\end{equation}

\begin{equation}
\frac{d\sigma_{\gamma Z}}{dr_T}=4\pi\alpha^2\varepsilon^2(g_V^2+g_A^2)\frac{1}{s\,r_T\sqrt{1-\dfrac{r_T^2}{r^2}}}\left(1-\dfrac{r_T^2}{2r^2}\right),\label{dist4}
\end{equation}
where

\begin{equation}
r=\frac{\sqrt{s}}{2m_Z\Gamma}.\label{radius}
\end{equation}
In a set of experimental data the fraction of the $Z'\rightarrow e^+e^-$ decay vertices located in the interval between $r_T$ and $r_T+dr_T$ from the beam axis is

\begin{equation}
f_{\gamma Z}(r_T)=\frac{1}{\sigma_{\gamma Z}}\frac{d\sigma_{\gamma Z}}{dr_T} \label{prob}
\end{equation}
with 

\begin{equation}
\sigma_{\gamma Z}=\int\limits_{r_{\text{m}}}^{r}\frac{d\sigma_{\gamma Z}}{dr_T}dr_T=\frac{4\pi\alpha^2\varepsilon^2(g_V^2+g_A^2)}{s}\left[\ln{\left(\frac{2r}{r_{\text{m}}}\right)}-\frac{1}{2}\right]. \label{prob1}
\end{equation}
Here $r_{\text{m}}$ is the minimal distance observable at the given experiment (it is assumed $r_{\text{m}}\ll r$). Then

\begin{equation}
f_{\gamma Z}(r_T)=\frac{1}{\left[\ln{\left(\dfrac{2r}{r_{\text{m}}}\right)}-\dfrac{1}{2}\right]r_T\sqrt{1-\dfrac{r_T^2}{r^2}}}\left(1-\dfrac{r_T^2}{2r^2}\right). \label{prob2}
\end{equation}
%\begin{equation}
%f_{\gamma Z}(r_T)=\frac{\left[\ln{\left(\dfrac{2r}{r_{\text{m}}}\right)}-\dfrac{1}{2}\right]^{-1}}{r_T\sqrt{1-\dfrac{r_T^2}{r^2}}}\left(1-\dfrac{r_T^2}{2r^2}\right). \label{prob2}
%\end{equation}
This distribution predicts a concentration of the decay vertices at $r_T\rightarrow r$ (see Fig.~\ref{fig4p}). Due to azimuthal symmetry such events will form a ring of radius $\sim r$ in the transverse plane with center at the interaction point of the beams, as schematically illustrated in Fig.~\ref{fig5p}.
Figure~\ref{fig6p} shows how the probability is distributed in this plane. An enhancement of the density of the $Z\rightarrow e^+e^-$ vertices at distances $r_T\simeq r$ is clearly seen.  Observation of such a ring can serve as a distinctive signature of the new bosons. In addition, measuring its radius $r$ allows to determine the product $m_Z\Gamma$, as it follows from equation~\eqref{radius}. Combining this with the number of vertices in the ring will give the mass of $Z'$ and its coupling strength separately. The width of the ring can be estimated by subtracting the distance corresponding to the minimum of~$f_{\gamma Z}(r_T)$ from $r$:

\begin{equation}
\text{ring width}\,\lesssim \left(1-\sqrt{\frac{2}{3}}\right)r. \label{width}
\end{equation}               

\begin{figure}
\includegraphics[width=1.1\textwidth]{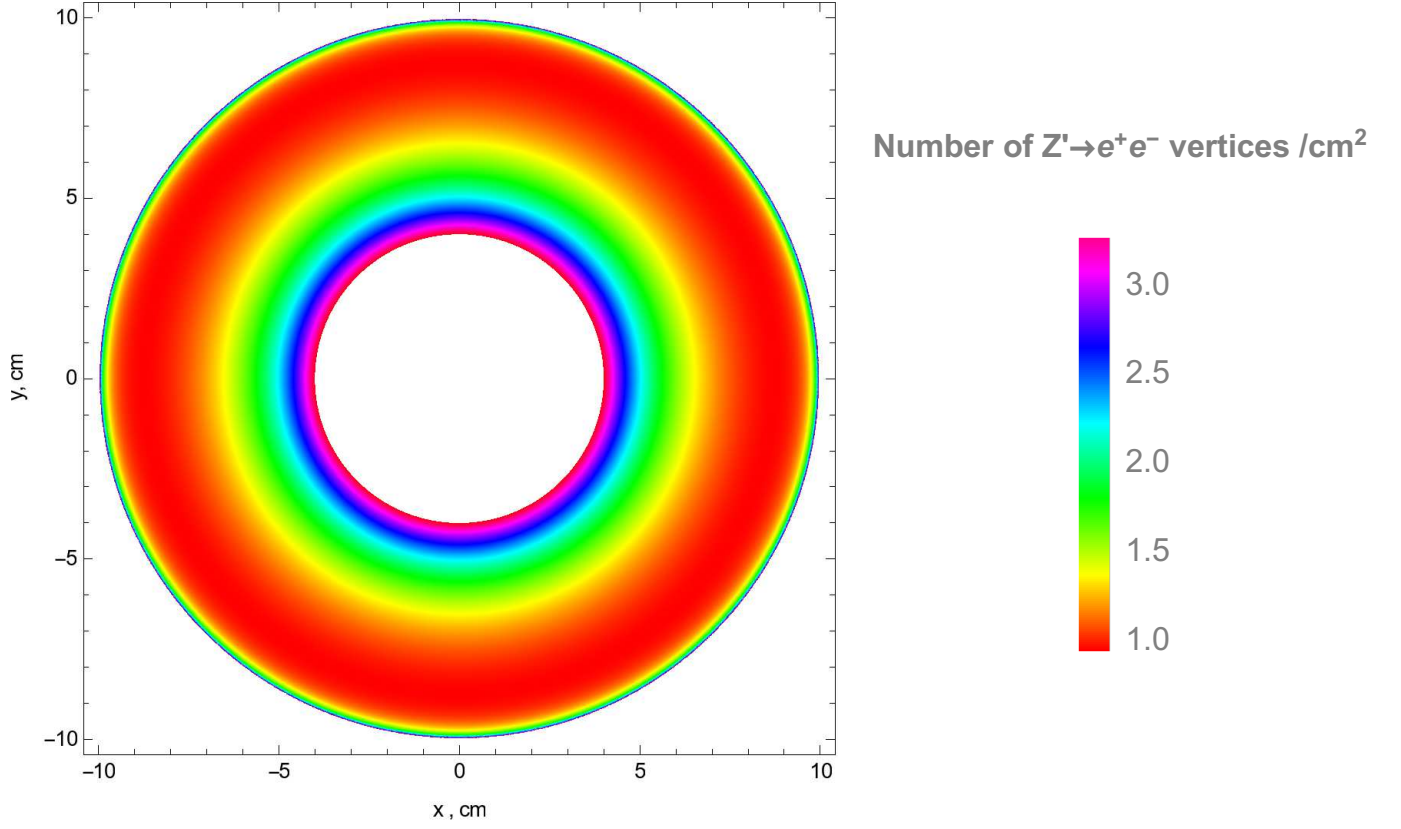}
\caption{Distribution of the $Z'\rightarrow e^+e^-$ vertices in the transverse plane corresponding to an integrated luminosity of $10^4$ fb$^{-1}$, $r_m=1$ cm, $r=10$ cm. The bosons are produced in $e^+e^-\rightarrow\gamma Z'$ at $\sqrt{s}=10.5$ GeV. The origin coincides with the interaction point of the colliding $e^+$ and $e^-$ beams. The region $\sqrt{x^2+y^2}\leq 4$ cm is cut out.}
\label{fig7p}
\end{figure}

\begin{figure}
\includegraphics[width=1.\textwidth]{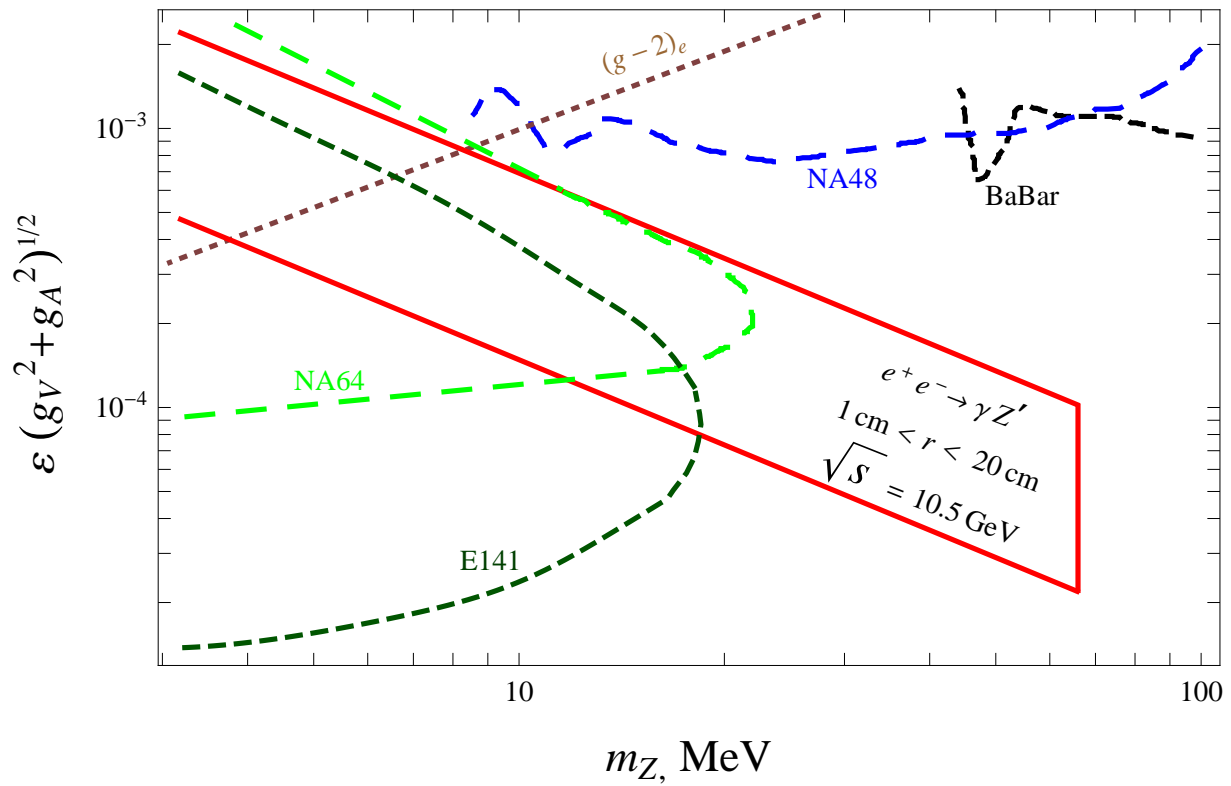}
\caption{Ranges for the coupling constant and  the mass of the $Z'$ accessible to collider experiments are located in the region between the red lines. They were computed for a center-of-mass energy $\sqrt{s}=10.5$ GeV and with the ability to identify  vertices for $Z'\rightarrow e^+e^-$ decays located at a distance $r$ between 1.0 and 20.0 cm from the beam axis. The range can be enlarged by increasing $r$ and/or the center-of-mass energy. We also include limits for the strength of the kinetic mixing between a new vector meson and the photon from experiments NA64~\cite{Banerjee:2018vgk}, NA48~\cite{Batley:2015lha}, BaBar~\cite{Lees:2014xha}, E141~\cite{Riordan:1987aw} and the electron anomalous magnetic moment $(g-2)_e$~\cite{Davoudiasl:2014kua}.}
\label{fig8p}
\end{figure}

In Fig.~\ref{fig7p} we plot a distribution of the vertices in the transverse plane for an integrated luminosity of $10^4$ fb$^{-1}$, $r_m=1$ cm and $r=10$~cm (the latter corresponds, for example, to $\sqrt{s}=10.5$ GeV, $\varepsilon(g_V^2+g_A^2)^{1/2}=1.2\times10^{-4}$, $m_Z=17$ MeV).                                                    
In this case the ring has the width about 1 cm and contains more than 10\% of the total number of vertices. Note that at a fixed cms energy different values of the parameters may lead to the same radius $r$. The ranges for the $Z'$ mass and coupling which are accessible to an experiment operating at $\sqrt{s}=10.5$ GeV and capable of measuring vertex positions at distances $1~\text{cm}< r< 20~\text{cm}$ from the beam axis are shown in Fig.~\ref{fig8p}. The above conditions overlap with the dimensions of the BaBar Silicon Vertex Tracker~\cite{Aubert:2001tu}. In the same figure we also include limits for the strength of the kinetic mixing between a new vector meson and the photon. The new vector meson is frequently introduced to mediate interactions between the dark and visible matter with the kinetic mixing inducing electromagnetic decays~\cite{Banerjee:2018vgk}. Finally, note that the parameters of the ring will be independent from the material of the detector because its formation is a consequence of the intrinsic properties of the boson (the $Z'$ lifetime). This would appear even in vacuum. At the same time, the positions of the background vertices from the $\gamma\rightarrow e^+e^-$ conversion processes will depend on the media since the related photon mean free path is proportional to the atomic number squared.

%%%%%%%%%%%%%%%%%%%%%%%%%%%%%%%%%%%%%%%%%%%%
\subsection{Energy spectrum of electrons in $e^+e^-\rightarrow\gamma (Z'\rightarrow e^+e^-)$}
%%%%%%%%%%%%%%%%%%%%%%%%%%%%%%%%%%%%%%%%%%%%
Special properties of the energy spectrum of the final electrons (or
positrons) coming from the boson decays in $e^+e^-\rightarrow\gamma (Z'\rightarrow e^+e^-)$ may serve as an additional signal for detecting the
new bosons. We compute it as

\begin{equation}
\frac{d\sigma_{\gamma Z}}{dE_e}=\int\frac{1}{\Gamma}\frac{d\Gamma}{dE_e}\frac{d\sigma_{\gamma Z}}{dE}dE\label{deed1},
\end{equation}
where $E_e$ is the final electron energy, $\dfrac{d\sigma_{\gamma Z}}{dE}$ is the cross section for the reaction
$e^+e^-\rightarrow\gamma Z'$ differential in the $Z'$ boson energy~$E$.
The latter equation, due to~\eqref{decayg6},  becomes 

\begin{equation}
\frac{d\sigma_{\gamma Z}}{dE_e}=\int\frac{1}{\sqrt{E^2-m_Z^2}}\frac{d\sigma_{\gamma Z}}{dE}dE,\label{deed22}
\end{equation}
in agreement, for example, with~\cite{Altarelli}.
Since $e^+e^-\rightarrow \gamma Z'$ is a $2\rightarrow2$ process, the
energy of the produced boson is always fixed in the cms reference frame, being $E=(s+m_Z^2-m_e^2)/2\sqrt{s}$. Therefore, neglecting the electron mass, we may write 

\begin{equation}
\frac{d\sigma_{\gamma Z}}{dE}=\sigma_{\gamma Z}\,\delta\left(E-\frac{s+m_Z^2}{2\sqrt{s}}\right),\label{deed23}
\end{equation}
where $\sigma_{\gamma Z}$ is the total cross section for $e^+e^-\rightarrow
\gamma Z'$, $\delta(x)$ is the delta function. The
$E$ integration in~\eqref{deed22} with \eqref{deed23} is trivial and we obtain 

\begin{equation}
\frac{1}{\sigma_{\gamma Z}}\frac{d\sigma_{\gamma Z}}{dE_e}=\frac{2\sqrt{s}}{s-m_Z^2}.\label{deed24}
\end{equation}
Thus, the energy spectrum of the final electrons is flat. Note that the spectrum of positrons will have exactly the
same behavior. This result is directly obtained in 
Appendix~\ref{app:decay}. 

\begin{figure}
\includegraphics[width=1.\textwidth]{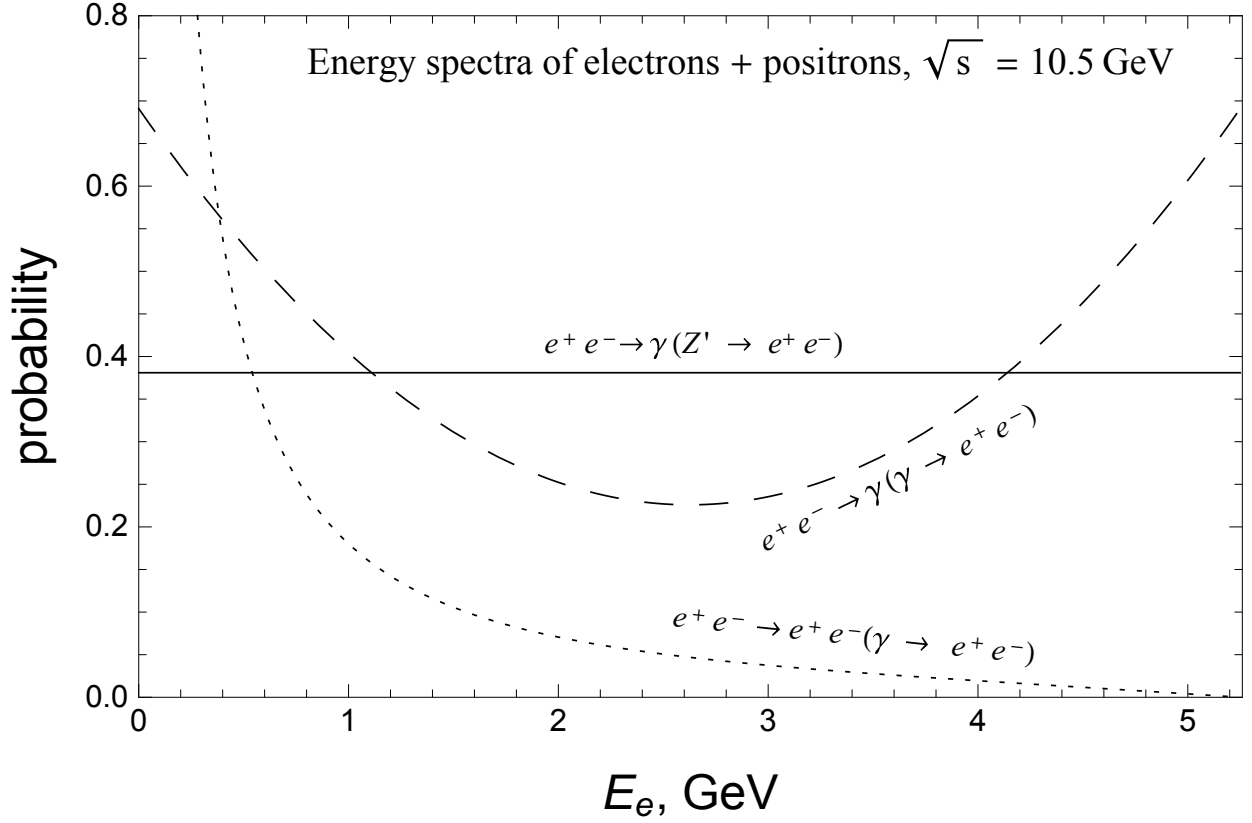}
\caption{Normalized energy spectra of electrons plus positrons produced in $e^+e^-$ collisions at $\sqrt{s}=10.5$ GeV. The horizontal solid line corresponds to an experimental data set containing the $Z'\rightarrow e^+e^-$ decay events only.  The dashed and dotted curves represent situations in which the data contain either only  $\gamma\rightarrow e^+e^-$ conversion events of photons from $e^+e^-\rightarrow\gamma\gamma$ or bremsstrahlung photon conversions in silicon, respectively.}
\label{fig9p}
\end{figure}

The main background to $e^+e^-\rightarrow\gamma (Z'\rightarrow e^+e^-)$ is expected from the processes $e^+e^-\rightarrow\gamma (\gamma\rightarrow e^+e^-)$ and $e^+e^-\rightarrow e^+e^-(\gamma\rightarrow e^+e^-)$ proceeding through the $\gamma\rightarrow e^+e^-$ conversion on atomic nuclei in the detector. In Fig.~\ref{fig9p} we compare the electron plus positron spectrum from  the $Z'$ decays  with that from the background at $\sqrt{s}=10.5$ GeV when the photons convert on silicon nuclei (we have chosen silicon because this is a frequently used element in particle detectors). As one can see, the spectra differ from each other qualitatively as well as quantitatively by a few tens of percent. An experimental data set of $\sim100$ electron events accumulated could be enough to resolve between the processes. It should be emphasized the picture presented alters very weakly from nucleus to nucleus.

%%%%%%%%%%%%%%%%%%%%%%%%%%%%%%%%%%%%%%%%%
\section{Vector bosons in $e^+e^-\rightarrow e^+e^-Z'$ \label{sectionm}}
%%%%%%%%%%%%%%%%%%%%%%%%%%%%%%%%%%%%%%%%%

Another channel of the production of the $Z'$ bosons, which plays an important role at certain kinematic conditions, is $e^+e^-\rightarrow e^+e^-Z'$.  Here we compute the corresponding cross sections and distributions of the observables introduced in the previous section. 

\subsection{Applicability of the equivalent photon approximation}

\begin{figure}
\includegraphics[width=1.\textwidth]{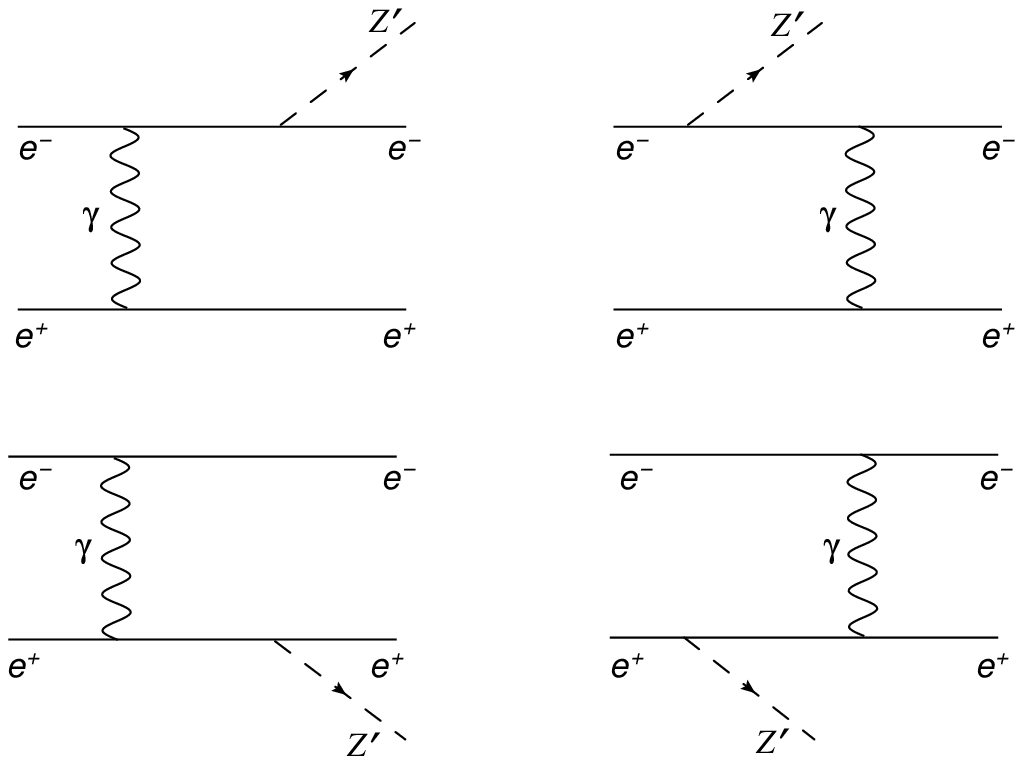}
\caption{Leading Feynman diagrams that contribute to $e^+e^-\rightarrow e^+e^- Z'$.}
\label{fig10p}
\end{figure}

Consider the reaction
\begin{equation}
e^+e^-\rightarrow e^+e^-Z'.\label{reac22n}
\end{equation}
The related leading Feynman diagrams are drawn in Fig.~\ref{fig10p}.
We will compute the cross sections in the Weizs\"acker-Williams equivalent photon approximation (EPA). According to EPA the reaction \eqref{reac22n} can be factorized into two subprocesses. The first one is emission of a photon by the electron (positron), $e\rightarrow e+\gamma$, the second is absorption of the emitted photon by the positron (electron) with the production of a $Z'$ boson:

\begin{equation}
e\gamma\rightarrow e Z'. \label{eq:reacsub}
\end{equation}
Then the total cross section of~\eqref{reac22n} can be represented as

\begin{equation}
\sigma(s)=2 \int\limits_{0}^{1}f(\eta,s)\hat\sigma(\eta s)\,d\eta,\label{dtt22}
\end{equation}
where $f(\eta,s)$ is the equivalent photon distribution of the electron (positron) with $\eta$ being the fraction of the electron (positron) energy carried by the photon, $\hat\sigma(s)$ is the total cross section for the subprocess~\eqref{eq:reacsub}. The factor 2 arises because the distributions for electrons and positrons coincide. Throughout this paper we adopt~(see Appendix~\ref{app:epa} for details)

\begin{equation}
f(\eta,s)=\frac{\alpha}{2\pi}\frac{1+(1-\eta)^2}{\eta}\ln{\left(\frac{\eta s^{3/2}}{m_Z m_e^2}\right)}.\label{dtts}
\end{equation}

This approximation is quite good for order of magnitude estimations, especially in view of the unknown coupling $\varepsilon$, which may vary over a wide range of values depending on a model. In addition, the interference between the two upper and two lower diagrams in Fig.~\ref{fig10p} is negligibly small in the limit $m_Z^2\ll s$. The point is that the $Z'$ boson will be predominantly emitted in the direction of the electron or positron so that the processes become distinguishable. Thus the cross section will be determined by the square of the sum of the two upper diagrams plus that of the lower ones. This is another justification for using EPA which considerably simplifies calculations.

\begin{figure}
\includegraphics[width=1.\textwidth]{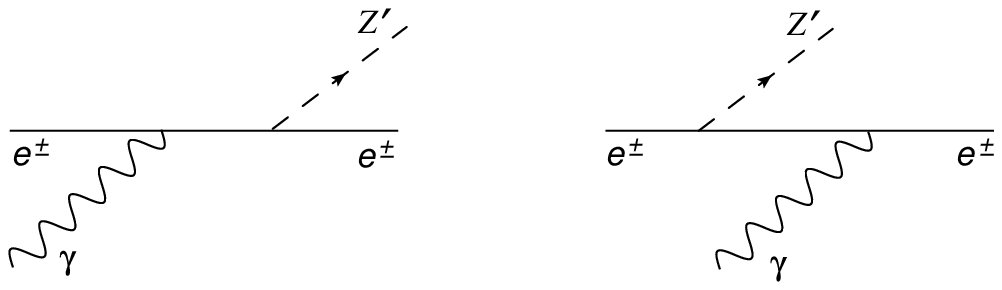}
\caption{Leading Feynman diagrams that contribute to subprocess $e\gamma\rightarrow e Z'$.}
\label{fig11p}
\end{figure}

\begin{figure}
\includegraphics[width=1.1\textwidth]{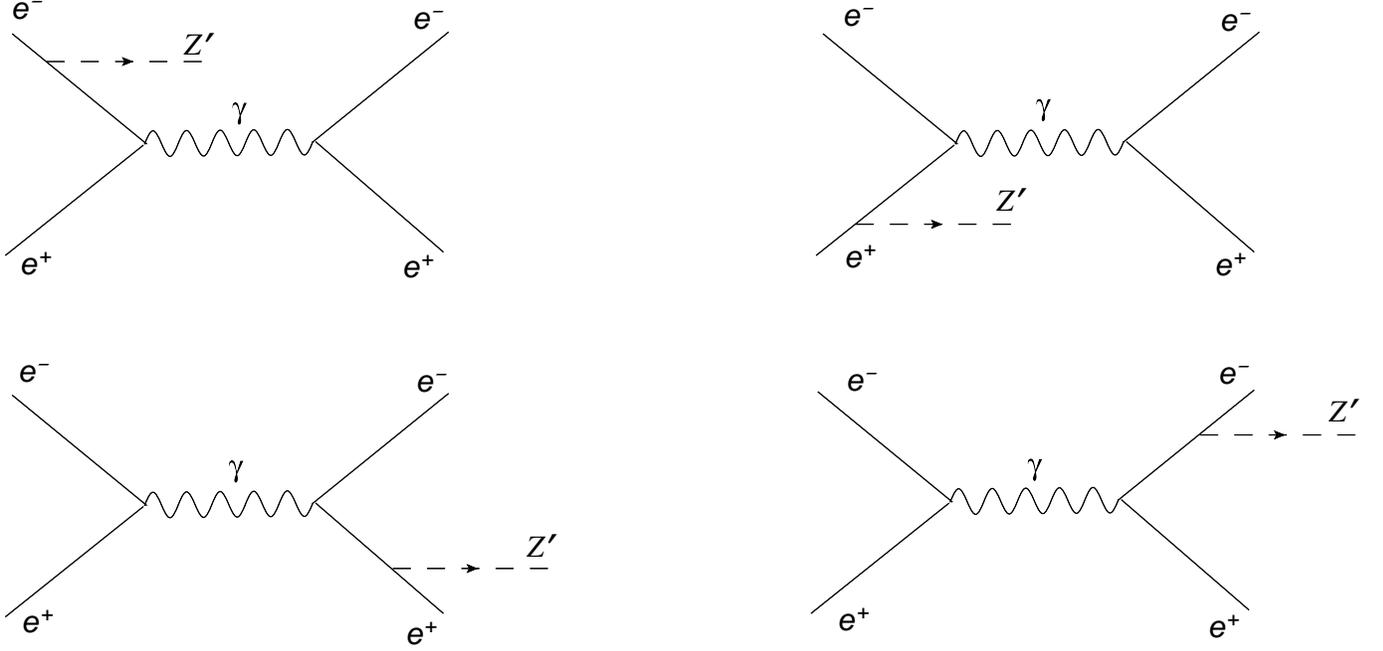}
\caption{Feynman diagrams with an $s$-channel photon exchange that contribute to $e^+e^-\rightarrow e^+e^- Z'$.}
\label{fig12p}
\end{figure}

The two lowest order Feynman diagrams contributing to~\eqref{eq:reacsub} are shown in Fig.~\ref{fig11p}. Noting that $e^-\gamma \rightarrow e^- Z'$ is related to the reaction $e^+e^-\rightarrow \gamma Z'$ by crossing symmetry we can immediately obtain the amplitude squared corresponding to~\eqref{eq:reacsub}. We need to exchange $t\leftrightarrow s$ in~\eqref{amplvl} so that

\begin{equation}
|M|^2(s,t,u)=-2 e^4 \varepsilon^2 (g_V^2+g_A^2) \left(\frac{u}{s}+\frac{s}{u}+\frac{2 m_Z^2t}{us}\right).\label{amplv}
\end{equation}
The overall sign changes because one fermion is crossed.
%After standard algebra one can obtain in the limit $s\gg m_Z^2\gg m_e^2$ that
%\begin{equation}
%\sigma(s)=\frac{4\alpha^3\varepsilon^2(g_V^2+g_A^2)}{m_Z^2}\ln{\left(\frac{m_Z^2}{m_e^2}\right)}\ln{\left(\frac{s}{m_e^2}\right)}.\label{sig2}
%\end{equation}
Note that full consideration would also require taking into account diagrams of the same order shown in Fig.~\ref{fig12p}. However they contain the exchange of an $s$-channel photon and produce terms $\sim1/s$ thus being greatly suppressed relative to the diagrams in Fig~\ref{fig10p}. This happens because the $t$-channel photon exchange in Fig~\ref{fig10p}  becomes highly dominating as the mass of the photon tends to zero.
This lead to an enhancement of the cross sections in models with light bosons due to the boson mass that appears in the denominators. These cross sections may be much larger than that for $e^+e^-\rightarrow \gamma Z'$. The main reason is that even though~$e^+e^-\rightarrow e^+e^-Z'$ is higher order in $\alpha$ compared to $e^+e^-\rightarrow \gamma Z'$, this suppression is compensated  by the soft photon exchange. 

Everywhere in the $\eta$-integrations of this section  we will keep only the leading terms.

  %This is clearly illustrated by the ratio of the cross sections in Fig.~\ref{fig4} at cms energies typical for experiments like BaBar~\cite{Aubert:2001tu} where the bosons can be searched. In Fig.~\ref{fig4}, we set $g_V=1$, $g_A=0$ and $m_Z=17$ MeV as hinted by the recent observations of the Atomki Collaboration~\cite{Krasznahorkay:2015iga}.  %The same property produces the dependence on the $Z'$ mass  shown in figure~\ref{fig5}, where one observes the significant dominance of reaction~\eqref{eq:reac2} as well.
%Thus, reaction~\eqref{eq:reac2} may be a promising channel for the production of new light gauge bosons that couple to electrons.

%%%%%%%%%%%%%%%%%%%%%%%%%%
\subsection{Energy spectrum of $Z'$ bosons}
\begin{figure}
\includegraphics[width=1.\textwidth]{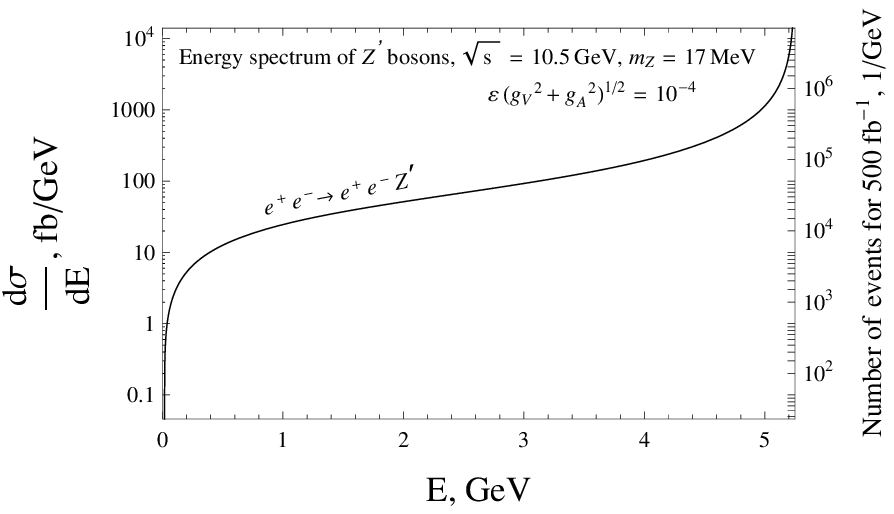}
\caption{Energy spectrum of  17 MeV vector bosons produced in $e^+e^-\rightarrow e^+e^-Z'$ at $\sqrt{s}=10.5$ GeV. The coupling is fixed to $\varepsilon(g_V^2+g_A^2)^{1/2}=10^{-4}$. The right-hand-side of the figure is labeled with the number of events corresponding to an integrated
luminosity of 500 fb$^{-1}$.}
\label{fig13p}
\end{figure}

%%%%%%%%%%%%%%%%%%%%%%%%%
Let us find out how the energies of the $Z'$ bosons produced in~$e^+e^-\rightarrow e^+e^-Z'$ will be distributed.
For this purpose we start with the cross section differential in the boson energy $E$:

\begin{equation}
\frac{d\sigma}{dE}=2 \int\limits_{\eta_{\text{min}}}^{\eta_{\text{max}}}f(\eta,s)\frac{d\hat\sigma}{dE}(\eta s)\,d\eta.\label{dtt}
\end{equation}
Here $\dfrac{d\hat\sigma}{dE}$ corresponds to the subprocess $e\gamma \rightarrow e Z'$ and, according to Appendix~\ref{app:pt}, can be written as
\begin{equation}
\frac{d\hat\sigma}{dE}(\eta s)=\frac{1}{8 \pi s^{3/2}}\frac{1}{\eta(1-\eta)}|M|^2(\eta s,t,u),\label{diff}
\end{equation}
where the amplitude squared is given by~\eqref{amplv}.
The integration limits are 

\begin{equation}
\eta_{\text{max}\atop \text{min}}=\frac{E}{\sqrt{s}}\left(1\pm\sqrt{1-\frac{m_Z^2}{E^2}}\right),\label{t_maxmin2}
\end{equation}
 as shown in Appendix~\ref{in_limit}. Using \eqref{dtt} with \eqref{diff}  we obtain

\begin{equation}
\frac{d\sigma}{dE}=32\alpha^3 \varepsilon^2(g_V^2+g_A^2)\frac{\sqrt{E^2-m_Z^2}}{m_Z^2s}\,\frac{2E^2-2E\sqrt{s}+s}{m_Z^2-2E\sqrt{s}+s}\ln\left(\frac{m_Z s}{2m_e^2E}\right),\label{dee}
\end{equation}
which determines the energy spectrum of the $Z'$ bosons. The $Z'$ energy may vary in the range 

\begin{equation}
m_Z\leq E\leq \frac{s+m_Z^2-m_e^2}{2\sqrt{s}}.\label{e_range}
\end{equation}
This spectrum is shown in Fig.~\ref{fig13p} for a 17 MeV boson with $\varepsilon(g_V^2+g_A^2)^{1/2}=10^{-4}$ produced at $\sqrt{s}=10.5$ GeV. One can see that most probably a boson carries a half of the total cms energy.

%%%%%%%%%%%%%%%%%%%%%%%%%%%%%%%%%%%%%%%%%%%%%%%%%
\subsection{The transverse momentum and the $Z'\rightarrow e^+e^-$ vertex position distributions \label{sec22}}
%%%%%%%%%%%%%%%%%%%%%%%%%%%%%%%%%%%%%%%%%%%%%%%%%%

In analogy with~\eqref{dtt} we may write 

\begin{equation}
\frac{d\sigma}{dp_T^2}=2 \int\limits_{4p_T^2/s}^1f(\eta,s)\frac{d\hat\sigma}{dp_T^2}(\eta s)\,d\eta,\label{dtt1}
\end{equation}
where $p_T$ is the transverse momentum of the $Z'$ boson. The subprocess cross section will read

\begin{equation}
\frac{d\hat\sigma}{dp_T^2}(\eta s)=\frac{1}{16\pi \eta^2 s^2\sqrt{1-\dfrac{4 p_T^2}{\eta s}}}|M|^2.\label{subpt}
\end{equation}
Compare the latter with~\eqref{subptf}. Using~\eqref{ptuts} in~\eqref{amplv} we obtain

\begin{equation}
|M|^2=-2 e^4\varepsilon^2 (g_V^2+g_A^2) \left(\frac{p_T^2}{t}+\frac{t}{p_T^2}+\frac{2 m_Z^2}{\eta^2 s^2} \frac{t^2}{p_T^2}\right).\label{amplvpt}
\end{equation}
The interference term can be omitted because it behaves as $\sim m_Z^2/p_T^2$ for $p_T\neq 0$ and vanishes at $p_T^2\rightarrow 0$ since

\begin{equation}
t=-\frac{\eta s}{2}\left(1-\sqrt{1-\frac{4 p_T^2}{\eta s}}\right). \label{tfpt}
\end{equation}
Therefore we may further write

\begin{equation}
|M|^2=-2e^4 \varepsilon^2 (g_V^2+g_A^2)\left(\frac{p_T^2}{t}+\frac{t}{p_T^2}\right).\label{amplvpt1}
\end{equation}
Substituting~\eqref{tfpt} into~\eqref{amplvpt1} we have performed the integration over $\eta$ in~\eqref{dtt1}  and arrived at the following transverse momentum   distribution of the $Z'$ bosons  in $e^+e^-\rightarrow e^+e^-Z'$:
%\begin{eqnarray}
%\frac{d\sigma}{dp_T^2}=\frac{\alpha^3\varepsilon^2(g_V^2+g_A^2)}{48 (p_T^2)^2}\left[8(7+2\beta)\sqrt{1-4 \beta}-3(7-8\beta-80\beta^2)\right.\nonumber\\\left.-48\beta\left[(2+\beta)\ln{\left(4\beta\right)}+(2-3\beta)\ln{\left(\frac{1-2\beta+\sqrt{1-4\beta}}{2\beta}\right)}\right]\right]\ln{\left(\frac{s}{m_e^2}\right)},\label{dptfin}
%\end{eqnarray}
%where $\beta=p_T^2/s$.  For events with $p_T^2\ll s$, equation~\eqref{dptfin} becomes compact:

\begin{equation}
\frac{d\sigma}{dp_T^2}=\frac{35\alpha^3\varepsilon^2(g_V^2+g_A^2)}{48 (p_T^2)^2}\ln{\left(\frac{4 \sqrt{s} p_T^2}{m_Zm_e^2}\right)}.\label{dptfin2}
\end{equation}
Note that this distribution very weakly depends on the boson mass. A plot of~\eqref{dptfin2} for the bosons with $\varepsilon(g_V^2+g_A^2)^{1/2}=10^{-4}$ produced at $\sqrt{s}=10.5$ GeV is shown in Fig.~\ref{fig2p}. The bosons are seen to have mostly small transverse momenta. Note that in this region the considered reaction significantly dominates over $e^+e^-\rightarrow \gamma Z'$.

Now we can also find out how the positions of the $Z'\rightarrow e^+e^-$ decay vertices will be distributed in the plane transverse to the colliding beams. Using~\eqref{dist2} with~\eqref{decayg5h} we rewrite~\eqref{dptfin2} in terms of $r_T$:

\begin{equation}
\frac{d\sigma}{dr_T}=\frac{105\alpha}{16 \varepsilon^2(g_V^2+g_A^2) m_Z^4 r_T^3}\ln{\left(\frac{4 m_Z\sqrt{s}\Gamma^2 r_T^2}{m_e^2}\right)}.\label{dptfin3}
\end{equation}  
Equation~\eqref{dptfin3} demonstrates that the vertices will be concentrated close to the beam axis. For example, the probability for a 17 MeV boson with $\varepsilon(g_V^2+g_A^2)^{1/2}\sim10^{-4}$ to decay at the distances of $\sim1$ cm  will be greatly suppressed because the denominator becomes as large as $\sim10^{27}$ MeV.

%%%%%%%%%%%%%%%%%%%%
\subsection{Angular distribution \label{subf}}
%%%%%%%%%%%%%%%%%%%%
Some care has to be taken when applying EPA to the derivation of the angular distribution of the $Z'$ bosons. 
Now the directions of motion of the equivalent photons from the incident electrons and positrons play a role.  The integral over $\eta$ must take into account that the $e^+$ and $e^-$ beams are opposite to each other. Therefore the cross section must be symmetrized as follows: 

%\begin{equation}
%\frac{d\sigma}{d\cos\theta}=\int\limits_{\frac{m_Z}{\sqrt{s}}}^1f_{\gamma/e}(\eta,s)\left[\frac{d\hat\sigma}{d\cos\theta}(\eta s,\cos\theta)+\frac{d\hat\sigma}{d\cos\theta}(\eta s,\cos(\pi-\theta))\right]\,d\eta.\label{dtt12}
%\end{equation}

\begin{equation}
\frac{d\sigma}{d\cos\theta}=\int\limits_{m_Z/\sqrt{s}}^1f(\eta,s)\left[\frac{d\hat\sigma}{d\cos\theta}(\eta s,\cos\theta)+\frac{d\hat\sigma}{d\cos\theta}(\eta s,\cos(\pi-\theta))\right]\,d\eta,\label{dtt12}
\end{equation}
%or 
%
%\begin{equation}
%\frac{d\sigma}{d\cos\theta}=\int\limits_{\sin^2\theta}^1f_{\gamma/e}(\eta,s)\left[\frac{d\hat\sigma}{d\cos\theta}(\eta s,\cos\theta)+\frac{d\hat\sigma}{d\cos\theta}(\eta s,-\cos\theta)\right]\,d\eta.\label{dtt13}
%\end{equation}
where $\theta$ is the scattering angle of $Z'$. The subprocess cross section standardly reads

\begin{equation}
\frac{d\hat\sigma}{d\cos\theta}(\eta s,\cos\theta)=\frac{1}{32\pi \eta s}|M|^2(\cos\theta).\label{dtt14}
\end{equation}
The squared amplitude can be obtained from~\eqref{amplvpt1} using $t=-\eta s(1-\cos\theta)/2$ and $u=-\eta s(1+\cos\theta)/2$: 

\begin{equation}
|M|^2(\cos\theta)=2 e^4\varepsilon^2 (g_V^2+g_A^2) \left(\frac{4+(1+\cos\theta)^2}{2 (1+\cos\theta)}\right).\label{dtt15}
\end{equation}
Thus we have

\begin{equation}
\frac{d\hat\sigma}{d\cos\theta}(\eta s,\cos\theta)+\frac{d\hat\sigma}{d\cos\theta}(\eta s,\cos(\pi-\theta))=\frac{4\pi\alpha^2\varepsilon^2(g_V^2+g_A^2)}{\eta s\sin^2\theta}\left(1+\frac{1}{4}\sin^2\theta\right). \label{dtt17}
\end{equation}
Finally,~\eqref{dtt12} with \eqref{dtt17} gives the angular distribution of the bosons  in $e^+e^-\rightarrow e^+e^-Z'$

\begin{equation}
\frac{d\sigma}{d\cos\theta}=\frac{4\alpha^3\varepsilon^2(g_V^2+g_A^2)}{m_Z\sqrt{s}\sin^2\theta}\left(1+\frac{1}{4}\sin^2\theta\right)\ln{\left(\frac{s}{m_e^2}\right)}.\label{angle}
\end{equation}

\begin{figure}
\includegraphics[width=1.\textwidth]{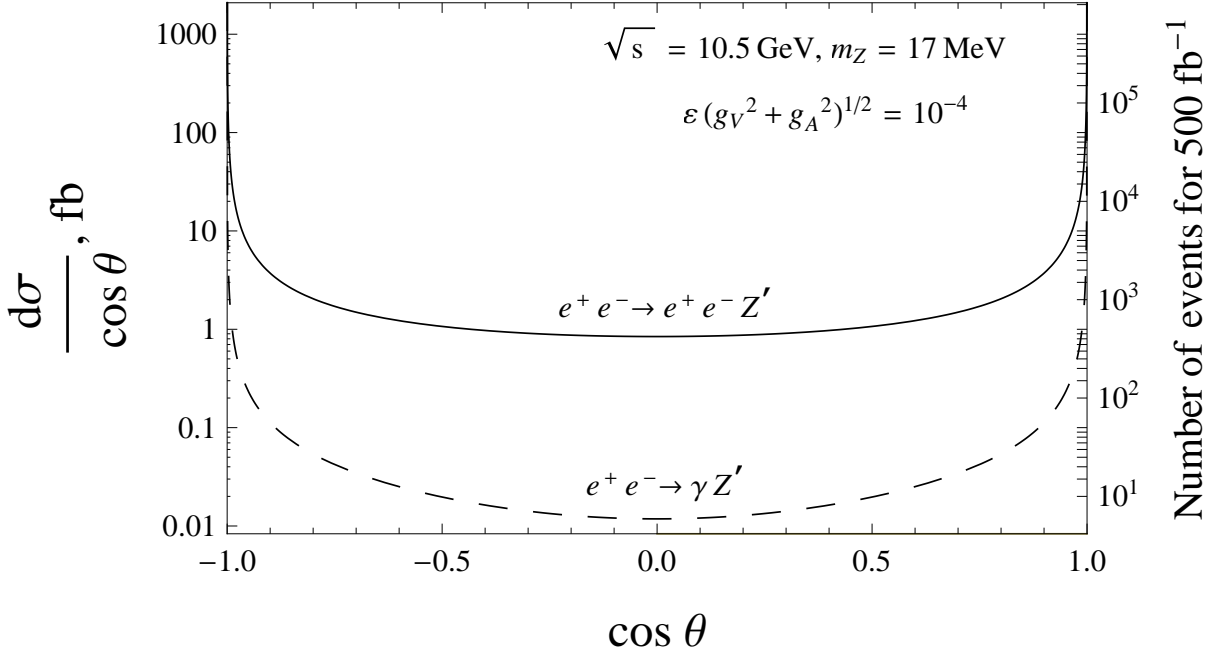}
\caption{Angular distribution of 17 MeV vector bosons with $\varepsilon(g_V^2+g_A^2)^{1/2}=10^{-4}$ produced in $e^+e^-\rightarrow e^+e^-Z'$ at $\sqrt{s}=10.5$ GeV (solid curve). The coupling is fixed to $\varepsilon(g_V^2+g_A^2)^{1/2}=10^{-4}$. For comparison the distribution in reaction $e^+e^-\rightarrow \gamma Z'$ given by~\eqref{angle13} at the same values of the parameters is also shown (dashed curve). The right-hand-side of the figure is labeled with the number of events corresponding to an integrated luminosity of 500 fb$^{-1}$.}
\label{fig14p}
\end{figure}

As an example, Fig.~\ref{fig14p} shows this distribution for a 17 MeV boson with $\varepsilon(g_V^2+g_A^2)^{1/2}=10^{-4}$ produced at $\sqrt{s}=10.5$ GeV. According to~\eqref{angle}, the bosons are predominantly scatter in the forward and backward directions, close to the beam axis. However, there is also a considerable fraction of events at large angles ($\cos\theta\simeq0$). Note that though the angles $\sim \pi/2$, such bosons, at the same time, carry small energies.

%%%%%%%%%%%%%%%%%%%%%%%%%%
\subsection{Energy spectrum of electrons in $e^+e^-\rightarrow e^+e^- (Z'\rightarrow e^+e^-)$}
%%%%%%%%%%%%%%%%%%%%%%%%%

\begin{figure}
\includegraphics[width=1.\textwidth]{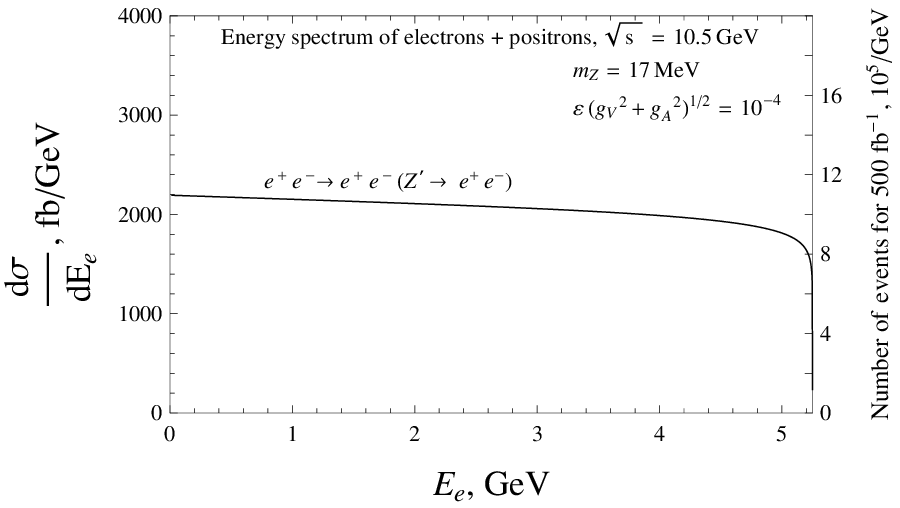}
\caption{Energy spectrum of electrons plus positrons from the decays of 17 MeV vector bosons produced in $e^+e^-\rightarrow e^+e^-Z'$ at $\sqrt{s}=10.5$ GeV. The coupling is fixed to $\varepsilon(g_V^2+g_A^2)^{1/2}=10^{-4}$. The right-hand-side of the figure is labeled with the number of events corresponding to an integrated luminosity of 500 fb$^{-1}$.}
\label{fig15p}
\end{figure}

As above, we compute the energy  spectrum of the electrons from the $Z'\rightarrow e^+e^-$ decays by substituting~\eqref{dee} into~\eqref{deed22}. The integration over the $Z'$ energy must be performed in the range

\begin{equation}
E_{\text{min}}=\frac{4E_e^2+m_Z^2}{4E_e},\,\,\,E_{\text{max}}=\frac{s+m_Z^2-m_e^2}{2\sqrt{s}}.\label{limits}
\end{equation}
The lower limit is a consequence of the condition $\cos\psi\leq1$ with $\psi$ being the angle between the three-momenta of $Z'$ and the outgoing electron. In contrast to $E_{\text{min}}$, where the electron mass can be safely neglected (as we have done), in $E_{\text{max}}$ the mass should be kept to ensure the regular behavior of the electron spectrum in the upper edge.
Thus we obtain the following energy spectrum of the final electrons:

%\begin{equation}
%\frac{d\sigma}{dE_e}=\frac{4 \alpha^3\varepsilon^2(g_V^2+g_A^2)}{m_Z^2s^{3/2}}\left[(\sqrt{s}-2E_e)^2+2s\ln\left(\frac{\left(\sqrt{s}-2E_e\right)(2E_e\sqrt{s}-m_Z^2)}{2E_e m_e^2}\right)\right]\ln\left(\frac{s}{m_e^2}\right).\label{distr_f}
%\end{equation} 

\begin{equation}
\frac{d\sigma}{dE_e}=\frac{8 \alpha^3\varepsilon^2(g_V^2+g_A^2)}{m_Z^2\sqrt{s}}\ln\left(\frac{\left(\sqrt{s}-2E_e\right)(2E_e\sqrt{s}-m_Z^2)}{2E_e m_e^2}\right)\ln\left(\frac{m_Z\sqrt{s}}{m_e^2}\right).\label{distr_f}
\end{equation} 
The corresponding expression for positrons is exactly the same.
We plot the spectrum of electrons plus positrons in Fig.~\ref{fig15p} for the case of a 17 MeV boson with $\varepsilon(g_V^2+g_A^2)^{1/2}=10^{-4}$ produced at $\sqrt{s}=10.5$ GeV.
For an integrated luminosity of 500 fb$^{-1}$ there may occur more than  $10^3$ $Z'\rightarrow e^+e^-$ decay events. Note that, similar to the reaction $e^+e^-\rightarrow \gamma Z'$, the spectrum~\eqref{distr_f} is also flat. This is a consequence of the fact that the  spectrum of $Z'$ bosons in $e^+e^-\rightarrow e^+e^- Z'$ peaks at the boson energies $\simeq \sqrt{s}/2$ exhibiting a delta function-like behavior (see Fig.~\ref{fig13p}).

\section{Light scalar bosons \label{sec2}}
%%%%%%%%%%%%%%%%%%%%%%%%%%%%%%%%%%%%%%%%%

We can extend the analysis and study the production of a neutral scalar boson (let us denote it by $\phi$) in  
\begin{equation}
e^+e^-\rightarrow \gamma\phi.\label{reacs}
\end{equation}
The interaction
\begin{equation}
\mathcal{L}_S=-eg\,\bar u_eu_e\phi\label{interact3}
\end{equation}
leads to the lowest order amplitude squared of the form

\begin{equation}
|M|_{S}^2(s,t,u)=e^4g^2\left(\frac{u}{t}+\frac{t}{u}+\frac{2m_{\phi}^2s}{ut}+2\right),\label{amplit2s}
\end{equation}
where $g$ is a Yukawa coupling.
One can proceed exactly as above and repeat all the calculations. 

It is important and relevant to compare~\eqref{amplit2s} with~\eqref{amplvl}. The dependences of both amplitudes  on the Mandelstam variables are identical except the last term  in~\eqref{amplit2s}. The latter will give a $\sim10\%$ contribution to the cross sections. In the situation when the coupling is unknown the observables will be described with a reasonable accuracy  even when this term is omitted in~\eqref{amplit2s}. This makes all the results of Sections~\ref{sec1d} and~\ref{sectionm} directly applicable to the scalar boson production when one makes the replacement

\begin{equation}
\varepsilon^2(g_V^2+g_A^2)\rightarrow \frac{g^2}{2},\,\,\,\,\,\,\,m_Z\rightarrow m_\phi.\label{amplit2}
\end{equation}
In other words, the cross sections as well as the distributions in this case will behave like those for the vector boson studied above in all details.

\section{Conclusions\label{concl}}
%%%%%%%%%%%%%%%%%
New light neutral gauge bosons beyond the Standard Model are of interest today as a possible explanation of several anomalies recently observed in experiments. These bosons may have non-zero couplings to electrons so that the most straightforward way to probe them is in $e^+e^-$ collisions.

In this paper, we have analyzed two reactions for the production of the new vector bosons with masses in the sub-GeV region. The first one is $e^+e^-\rightarrow\gamma Z'$. The second one, $e^+e^-\rightarrow e^+e^-Z'$, we have treated within the framework of the equivalent photon approximation. Analytic expressions for the cross sections and the energy spectra of the produced bosons as well as of the final electrons from the boson decays are obtained. In particular, we have studied the transverse momentum distribution of the bosons and the  distribution of the positions of $Z'\rightarrow e^+e^-$ decay vertices with respect to the axis of the colliding beams. It is shown that these distributions can serve as distinct signatures of the bosons in $e^+e^-\rightarrow\gamma Z'$. The reaction $e^+e^-\rightarrow e^+e^-Z'$ becomes important at small $Z'$ scattering angles where its contribution to the overall yield may be larger by orders of magnitude compared to $e^+e^-\rightarrow\gamma Z'$. 
Numerical predictions, at energies $\sqrt{s}=10.5$ GeV, typical for the B-factories are derived.

The dominant background to the considered reactions is expected from the $\gamma\rightarrow e^+e^-$ conversion of photons appearing from $e^+e^-\rightarrow \gamma\gamma$ and the QED bremsstrahlung. These $e^+e^-$ pairs can be disentangled from the $Z'\rightarrow e^+e^-$ decays by measuring the energy spectra of the final electrons (positrons) and/or the spatial distribution the $e^+e^-$ vertices.  As to the direct production of $e^+e^-$ pairs, $e^+e^-\rightarrow e^+e^-e^+e^-$, selecting the vertices clearly separated from the collision point of the beams~\cite{Bjorken:2009mm,Bossi:2013lxa} reduces this background. 

We have also discussed the possibility of the boson decay into neutrino--antineutrino pairs. In this case the boson can manifest itself as missing transverse momentum. The production of a scalar boson is also investigated.
The calculations are performed in full detail and can be useful for similar studies.

%%%%%%%%%%%%%%%%%%%%%%%%
\begin{acknowledgments}
One of us (I.~A.) was supported in part by the Program of fundamental scientific research of the Presidium of the Russian Academy of Sciences "Physics of fundamental interactions and nuclear technologies".
\end{acknowledgments}

\appendix
  
%%%%%%%%%%%%%%%%%%%%%%%%%%%%%%%%%%
\section{Kinematics and cross sections\label{app:pt}}
%%%%%%%%%%%%%%%%%%%%%%%%%%%%%%%%%%
In order to find $\dfrac{d\hat\sigma}{dE}$ which appears in \eqref{dtt} we begin with 

\begin{equation}
\frac{d\hat\sigma}{dt}=\frac{1}{16\pi \eta^2 s^2}|M|^2(\eta s,t,u).\label{dsigt}
\end{equation}
This is a well known textbook formula when $\eta=1$.
We choose the z-axes to point in the electron beam direction.
Then, by definition, for the subprocess $e^+\gamma\rightarrow e^+Z'$ 

\begin{equation}
u=(p_e'-p_Z)^2=m_e^2+m_Z^2-\sqrt{s}(E+p_{L}).\label{for_u}
\end{equation}
Here  $p_e'$ is the incident positron four-momentum, $E$ and $p_L$ are the energy and longitudinal momentum of $Z'$, respectively. We have also used that the positron energy  is $\sqrt{s}/2$.
Likewise

\begin{equation}
t=(p_\gamma-p_Z)^2=m_Z^2-\eta\sqrt{s}(E-p_L).\label{for_t}
\end{equation}
Note that $E_\gamma=\eta E_e=\eta\sqrt{s}/2$.
On the other hand, there is also the condition

\begin{equation}
\eta s+t+u=2m_e^2+m_Z^2.\label{cond}
\end{equation}
Adding \eqref{for_u} to~\eqref{for_t} and using~\eqref{cond} yield

\begin{equation}
p_L=\frac{m_Z^2-m_e^2-E\sqrt{s}(1+\eta)+\eta s}{\sqrt{s}(1-\eta)}.\label{for_pl1}
\end{equation}
Substituting~\eqref{for_pl1} into \eqref{for_t} we find the relation between $t$ and $E$

\begin{equation}
t=\frac{m_Z^2-\eta(m_e^2+2E\sqrt{s}-\eta s)}{1-\eta}.\label{for_pl}
\end{equation}
Since

%\begin{equation}
%\left|\frac{dt}{dE}\right|=2\sqrt{s}\frac{\eta}{1-\eta}
%\end{equation}
%and

\begin{equation}
\frac{d\hat\sigma}{dE}=\frac{d\hat\sigma}{dt}\left|\frac{dt}{dE}\right|.
\end{equation}
we obtain

\begin{equation}
\frac{d\hat\sigma}{dE}=\frac{1}{8 \pi s^{3/2}}\frac{1}{\eta(1-\eta)}|M|^2(\eta s,t,u), \label{diffap}
\end{equation}
which is in agreement, for example, with equation (53) of~\cite{Altarelli}.

In analogy, if we know the relation between $t$ and $p_T^2$, then

\begin{equation}
\frac{d\hat\sigma}{dp_T^2}=\frac{d\hat\sigma}{dt}\left|\frac{dt}{dp_T^2}\right|.\label{pt_cross}
\end{equation}
Neglect the square of the electron mass in~\eqref{for_u} and write down the following product

\begin{equation}
(u-m_Z^2)(t-m_Z^2)=\eta s (E^2-p_L^2)=\eta s p_T^2+\eta s m_Z^2,\label{ut}
\end{equation}
where  we have used $E^2=p_L^2+p_T^2+m_Z^2$. 
On the other hand
\begin{eqnarray}
(u-m_Z^2)(t-m_Z^2) =ut -m_Z^2(u+t)+m_Z^4= ut -m_Z^2(m_Z^2-\eta s)+m_Z^4 \nonumber\\=ut+\eta s m_Z^2.\label{ut1}
\end{eqnarray}
Comparing~\eqref{ut} with~\eqref{ut1} we obtain

\begin{equation}
p_T^2=\frac{ut}{\eta s}\label{utsapp}.
\end{equation}
Using~\eqref{cond} rewrite~\eqref{utsapp} in the form

\begin{eqnarray}
p_T^2=\frac{(m_Z^2-\eta s-t)t}{\eta s}\label{utsfapp}.
\end{eqnarray}
Now we can solve~\eqref{utsfapp} as an equation with respect to $t$. The kinematically allowed solution reads

\begin{equation}
t=-\frac{\eta s-m_Z^2}{2}\left(1-\sqrt{1-\frac{4\eta s p_T^2}{(\eta s-m_Z^2)^2}}\right).
\end{equation}
If the problem also justifies neglecting $m_Z^2$ then

\begin{equation}
t=-\frac{\eta s}{2}\left(1-\sqrt{1-\frac{4 p_T^2}{\eta s}}\right), \label{tfptapp}
\end{equation}

\begin{equation}
\left|\frac{dt}{dp_T^2}\right|=\frac{1}{\sqrt{1-\dfrac{4 p_T^2}{\eta s}}}
\end{equation}
and according to~\eqref{pt_cross}

\begin{equation}
\frac{d\hat\sigma}{dp_T^2}=\frac{1}{16\pi \eta^2 s^2\sqrt{1-\dfrac{4 p_T^2}{\eta s}}}|M|^2.\label{subpt23}
\end{equation}
Equations~\eqref{utsapp}--\eqref{subpt23} apparently hold at $\eta=1$, so that we can check the correctness of~\eqref{subptf2} as follows. By definition $p_T^2=\dfrac{s}{4}\sin^2\theta=\dfrac{s}{4}(1-\cos^2\theta)$, which leads to 

\begin{equation}
\left|\frac{dp_T^2}{d\cos\theta}\right|=\frac{s}{2}|\cos\theta| \,\,\,\text{with}\,\,\,|\cos\theta|=\sqrt{1-\frac{4p_T^2}{s}}. 
\end{equation}
Therefore

\begin{eqnarray}
\frac{d\sigma_{\gamma Z}}{d\cos\theta}=\frac{d\sigma_{\gamma Z}}{dp_T^2}\left|\frac{dp_T^2}{d\cos\theta}\right|=\frac{4\pi(g_V^2+g_A^2)\varepsilon^2\alpha^2}{s \sin^2\theta}\left(1-\frac{1}{2}\sin^2\theta\right).\label{angle13}
\end{eqnarray}
This is the angular distribution of the vector bosons in $e^+e^-\rightarrow\gamma Z'$~\cite{Fayet:2007ua,Borodatchenkova:2005ct,Chen:2016dhm}.

%%%%%%%%%%%%%%%%%%%%%%%%%%%%%%%%%%%%%%%%%%
\section{The boson decays\label{app:decay}}
%%%%%%%%%%%%%%%%%%%%%%%%%%%%%%%%%%%%%%%%

Usually a decay of a particle is considered in its rest reference frame.
Meanwhile, there are problems in which the particles are produced
with nonzero three-momenta and one has to analyze the energy spectra
of their decay products. Such a situation arises, for example, in the
process $e^+e^-\rightarrow\gamma (Z'\rightarrow e^+e^-)$. Let us derive the energy
spectrum of electrons coming from the in-flight decays of the unpolarized vector bosons

\begin{eqnarray}
Z'(p_Z)\rightarrow e^+(p_e')e^-(p_e),
\end{eqnarray}
where the four-momenta are given in the parentheses, $p_Z=(E,
\mathbf{p})$, $p_e=(E_e, \mathbf{q_e})$.
For the interaction in~\eqref{interact2}  the square of the
matrix element summed over spin states and averaged over initial polarizations is
\begin{equation}
|M|^2(p_Z,p_e,p_e')=\frac{8\pi\alpha\varepsilon^2(g_V^2+g_A^2)}{3m_Z^2}
\left[4(p_Zp_e)(p_Zp_e')+m_Z^4\right].\label{ampld}
\end{equation}
The electron mass is neglected. Following standard rules we obtain
%The decay width is given by

%\begin{equation}
%d\Gamma=\frac{1}{(2\pi)^2}\frac{1}{2E}|M|^2(p_Z,p_e,p_e')\delta^{4}(p_Z-p_e-p_e')\frac{d^3\mathbf{q_e}}{2E_e}\frac{d^3\mathbf{q'_e}}{2E_e'}.\label{decayg}
%\end{equation}
%Taking into account the identity

%\begin{equation}
%\frac{d^3\mathbf{q'_e}}{2E_e'}=\Theta(E_e')\delta(p_e'^2-m_e^2)d^4p'_e,\label{identity}
%\end{equation}
%where $\Theta(E_e')$ is the step function, we can
%integrate~\eqref{decayg} over $p_e'$ exploiting the four-dimensional
%delta function. Then

%\begin{equation}
%d\Gamma=\frac{1}{(2\pi)^2}\frac{1}{2E}|M|^2(p_Z,p_e,p_Z-p_e)\delta((p_Z-p_e)^2-m_e^2)\frac{d^3\mathbf{q_e}}{2E_e}.\label{decayg1}
%\end{equation}
%On the other hand

%\begin{equation}
%(p_Z-p_e)^2\simeq
%m_e^2+m_Z^2-2EE_e\left(1-\frac{p}{E}\cos\psi\right)=m_e^2+m_Z^2-2xEE_e,\label{app1}
%\end{equation}
%where $x=\left(1-\dfrac{p}{E}\cos\psi\right)$ with
%$\psi$ being the angle between $\mathbf{p}$ and $\mathbf{q_e}$.  Also
%\begin{equation}
%\frac{d^3\mathbf{q_e}}{2E_e}\simeq\frac{\pi
%EE_e}{p}dq_edx.\label{app2}
%\end{equation}
%The sign "$\simeq$" is used because in typical collision experiments the observed electron energy $E_e\gg m_e$ and therefore
%$E_e\simeq q_e$.
%Putting all these things together, we have

%\begin{equation}
%d\Gamma=\frac{\alpha\varepsilon^2(g_V^2+g_A^2)}{6m_Z^2pE}\left[4xEE_e(m_Z^2-xEE_e)+m_Z^4\right]\delta\left(x-\frac{m_Z^2}{2EE_e}\right)dq_edx.\label{decayg3}
%\end{equation}
%The $x$ integration in~\eqref{decayg3} is trivial due to the
%delta function:

\begin{equation}
\frac{d\Gamma}{dE_e}=\frac{\alpha\varepsilon^2(g_V^2+g_A^2)}{3pE}m_Z^2\label{decayg4}
\end{equation}
and

\begin{equation}
\Gamma=\frac{\alpha\varepsilon^2(g_V^2+g_A^2)}{3\gamma(v)}m_Z.\label{decayg5}
\end{equation}
Here we emphasize the presence of the Lorentz factor, $\gamma(v)=E/m_Z$, in the denominator. For the decay at rest $\gamma(v)=1$.
Dividing~\eqref{decayg4} by~\eqref{decayg5} we finally obtain the
spectrum of electrons coming from the decays of $Z'$ bosons
of given momentum $\mathbf{p}$:

\begin{equation}
\frac{1}{\Gamma}\frac{d\Gamma}{dE_e}=\frac{1}{p}.\label{decayg6}
\end{equation}
The spectrum is flat, independent on the electron momentum.
For example, if the reaction~$e^+e^-\rightarrow\gamma Z'$ proceeds at
the center-of-mass energy $\sqrt{s}$, then $p=(s-m_Z^2)/{2\sqrt{s}}$ and
the final electron energies will be distributed as

\begin{equation}
\frac{1}{\Gamma}\frac{d\Gamma}{dE_e}=\frac{2\sqrt{s}}{s-m_Z^2}.\label{decayg7}
\end{equation}
The spectrum of positrons from the~$Z'\rightarrow e^+e^-$ decays is the
same since~\eqref{ampld} is symmetric under the exchange
$p_e\leftrightarrow p_e'$.

If we consider a similar decay mode of the scalar boson $\phi(p_{\phi})\rightarrow e^+(p_e')e^-(p_e)$ with the interaction~\eqref{interact3}, then

\begin{eqnarray}
|M|^2(p_\phi,p_e,p_e')=8\pi\alpha g^2 m_\phi^2.\label{amplds2}
\end{eqnarray}
One can show that the related partial width in the boson rest frame is
\begin{equation}
\Gamma_S=\frac{\alpha g^2}{2}m_\phi.\label{decayg5ss}
\end{equation}
The energy spectrum of the outgoing electrons (positrons) exactly repeats the one for $Z'$ given by~\eqref{decayg7}.

%%%%%%%%%%%%%%%%%%%%%%%%%%%		
\section{The $\eta$-integration limits \label{in_limit}}
%%%%%%%%%%%%%%%%%%%%%%%%%%%
Here we derive the limits of integration over $\eta$ in the calculations of the angular and energy distributions of the $Z'$ bosons produced in the reaction $e^+e^-\rightarrow e^+e^-Z'$.
Neglecting the square of the electron mass, rewrite~\eqref{for_pl1} as

%\begin{equation}
%p_L=\frac{m_Z^2-m_e^2-E\sqrt{s}(1+\eta)+\eta s}{\sqrt{s}(1-\eta)}.\label{for_pl1pp}
%\end{equation} 
%Neglecting the square of the electron mass, from~\eqref{for_pl1pp} one finds

\begin{equation}
\eta=-\frac{E+p_L-\dfrac{m_Z^2}{\sqrt{s}}}{E-p_L-\sqrt{s}}.\label{for_pl1pq}
\end{equation} 
Then using $p_L=\sqrt{E^2-m_Z^2}\, \cos{\theta}$ we obtain

\begin{equation}
\eta=-\frac{E+\sqrt{E^2-m_Z^2}\, \cos{\theta}-\dfrac{m_Z^2}{\sqrt{s}}}{E-\sqrt{E^2-m_Z^2}\, \cos{\theta}-\sqrt{s}}.\label{for_pl1pq1}
\end{equation}
Thus, we have $\eta$ as a function of two independent variables, $E$ and $\theta$.

In order to find the integration limits in the case of the angular distribution, we fix the angle $\theta$ and vary the boson energy in the interval 

\begin{equation}
E_{\text{min}}=m_Z,\,\,\,E_{\text{max}}=\frac{s+m_Z^2}{2\sqrt{s}}.\label{limitsnn}
\end{equation}
We find from~\eqref{for_pl1pq1} that the corresponding $\eta$ limits are
\begin{equation}
\eta_{\text{min}}=\frac{m_Z}{\sqrt{s}},\,\,\,\eta_{\text{max}}=1.\label{limits_eta}
\end{equation}

In the case of the energy distribution, we fix the energy $E$ and vary the longitudinal momentum (in other words  the angle $\theta$). Obviously,  $p_{L\,\text{min}}=-\sqrt{E^2-m_Z^2}$ and $p_{L\,\text{max}}=\sqrt{E^2-m_Z^2}$, so that from~\eqref{for_pl1pq} we obtain

\begin{equation}
\eta_{\text{max} \atop \text{min}}=\frac{E}{\sqrt{s}}\left(1\pm\sqrt{1-\frac{m_Z^2}{E^2}}\right). \label{t_maxmin212}
\end{equation}

\section{Scale dependence of the equivalent photon distribution\label{app:epa}}
%%%%%%%%%%%%%%%%%%%%%%%%%%%
Here we derive the scale dependence of the equivalent photon distribution of the electron which can be represented as~\cite{Altarelli,Chen:1975sh}:

\begin{equation}
f(\eta,s)=\frac{\alpha}{2\pi}\frac{1+(1-\eta)^2}{\eta}\ln{\left(\frac{Q^2_{\text{max}}}{Q^2_{\text{min}}}\right)}.\label{app1d}
\end{equation}
Note that $Q^2=-t$. Then, using~\eqref{for_pl}
 at fixed $\eta$ we can write

\begin{equation}
Q^2_{\text{max}}=-\frac{m_Z^2-\eta(m_e^2+2E_{\text{max}}\sqrt{s}-\eta
s)}{1-\eta}.\label{for_pl23456}
\end{equation}
Since $$E_{\text{max}}=\dfrac{s+m_Z^2-m_e^2}{2\sqrt{s}}$$ one obtains

\begin{equation}
Q^2_{\text{max}}=\eta s-m_Z^2 .\label{for_pl234567}
\end{equation}
In analogy

\begin{equation}
Q^2_{\text{min}}=-\frac{m_Z^2-\eta(m_e^2+2E_{\text{min}}\sqrt{s}-\eta
s)}{1-\eta}.\label{for_pl2345678}
\end{equation}
Obviously, $E_{\text{min}}=m_Z$. However, there is the following subtlety in this case. Equation~\eqref{for_pl1pq1} tells us that only one value of $\eta$ corresponds to $E=m_Z$, namely $\eta=m_Z/\sqrt{s}$. Therefore,~\eqref{for_pl2345678} becomes

\begin{equation}
Q^2_{\text{min}}=m_e^2 \frac{m_Z}{\sqrt{s}-m_Z}.\label{for_pl234567890}
\end{equation}
Finally, substituting~\eqref{for_pl234567} and~\eqref{for_pl234567890} into~\eqref{app1d} we find

\begin{equation}
f(\eta,s)=\frac{\alpha}{2\pi}\frac{1+(1-\eta)^2}{\eta}\ln{\left(\frac{(\eta
s-m_Z^2)(\sqrt{s}-m_Z)}{m_Zm_e^2}\right)}.\label{app1dfin4}
\end{equation}
As long as $m_Z^2\ll \eta s$

\begin{equation}
f(\eta,s)=\frac{\alpha}{2\pi}\frac{1+(1-\eta)^2}{\eta}\ln{\left(\frac{\eta
s^{3/2}}{m_Zm_e^2}\right)}.
\end{equation}

%%%%%%%%%%%%%%%%%%%%%%%%%%%%%%%%
%%References
%%%%%%%%%%%%%%%%%%%%%%%%%%%%%%%%

\newpage
%%%%%%%%
%%%%%%%%%%%%%%% figures
%%%%%%%%

%%%%%%%%%%%%%%
% FIGURE 1
%%%%%%%%%%%%%%

%%%%%%%%%%%%%%
% FIGURE 2
%%%%%%%%%%%%%%

%%%%%%%%%%%%%%
% FIGURE 3
%%%%%%%%%%%%%%

%%%%%%%%%%%%%%
% FIGURE 4
%%%%%%%%%%%%%%

%%%%%%%%%%%%%%
% FIGURE 5
%%%%%%%%%%%%%%

%%%%%%%%%%%%%%
% FIGURE 6
%%%%%%%%%%%%%%

%%%%%%%%%%%%%%
% FIGURE 7
%%%%%%%%%%%%%%

%%%%%%%%%%%%%%
% FIGURE 8
%%%%%%%%%%%%%%

%%%%%%%%%%%%%%
%FIGURE 9
%%%%%%%%%%%%%%

%%%%%%%%%%%%%%
% FIGURE 10
%%%%%%%%%%%%%%

%%%%%%%%%%%%%%
% FIGURE 11
%%%%%%%%%%%%%%

%%%%%%%%%%%%%%
% FIGURE 12
%%%%%%%%%%%%%%

%%%%%%%%%%%%%%
% FIGURE 13
%%%%%%%%%%%%%%

%%%%%%%%%%%%%%
% FIGURE 14
%%%%%%%%%%%%%%

%%%%%%%%%%%%%%
% FIGURE 15
%%%%%%%%%%%%%%

%%%%%%


\begin{thebibliography}{99}


%\cite{Langacker:1980js}
\bibitem{Langacker:1980js} 
  P.~Langacker,
  %``Grand Unified Theories and Proton Decay,''
  Phys.\ Rept.\  {\bf 72}, 185 (1981).

\bibitem{Hewett:1988xc} 
  J.~L.~Hewett and T.~G.~Rizzo,
  %``Low-Energy Phenomenology of Superstring Inspired E(6) Models,''
  Phys.\ Rept.\  {\bf 183}, 193 (1989).

%\cite{Langacker:2008yv}
\bibitem{Langacker:2008yv} 
  P.~Langacker,
  %``The Physics of Heavy $Z^\prime$ Gauge Bosons,''
  Rev.\ Mod.\ Phys.\  {\bf 81}, 1199 (2009)
   [arXiv:0801.1345 [hep-ph]].

\bibitem{Altarelli}
G.~Altarelli, G.~Martinelli, B.~Mele and R.~R\"uckl,
  %``Electroproduction of Supersymmetric Particles and Gauge Bosons,''
  Nucl.\ Phys.\ B {\bf 262}, 204 (1985).

\bibitem{Aliev}
T.~M.~Aliev and A.~A.~Bayramov, 
  Sov.\ J.\ Nucl.\ Phys.\ {\bf 52}, 689 (1990).

%\cite{Aad:2015osa}
\bibitem{Aad:2015osa} 
  G.~Aad {\it et al.} [ATLAS Collaboration],
  %``A search for high-mass resonances decaying to $\tau^{+}\tau^{-}$ in $pp$ collisions at $\sqrt{s}=8$ TeV with the ATLAS detector,''
  JHEP {\bf 1507}, 157 (2015)
  [arXiv:1502.07177 [hep-ex]].

%\cite{CMS:2016zxk}
\bibitem{CMS:2016zxk} 
  CMS Collaboration [CMS Collaboration],
  %``Search for new physics with high-mass tau lepton pairs in pp collisions at sqrt(s) = 13 TeV with the CMS detector,''
  CMS-PAS-EXO-16-008.

%\cite{ATLAS:2016cyf}
\bibitem{ATLAS:2016cyf} 
  The ATLAS collaboration [ATLAS Collaboration],
  %``Search for new high-mass resonances in the dilepton final state using proton-proton collisions at $\sqrt{s}$ = 13 TeV with the ATLAS detector,''
  ATLAS-CONF-2016-045.

%\cite{CMS:2016abv}
\bibitem{CMS:2016abv} 
  CMS Collaboration [CMS Collaboration],
  %``Search for a high-mass resonance decaying into a dilepton final state in 13 fb$^{-1}$ of pp collisions at $\sqrt{s}=13~\mathrm{TeV}$,''
  CMS-PAS-EXO-16-031.

%\cite{Erler:2009jh}
\bibitem{Erler:2009jh} 
  J.~Erler, P.~Langacker, S.~Munir and E.~Rojas,
  %``Improved Constraints on Z-prime Bosons from Electroweak Precision Data,''
  JHEP {\bf 0908}, 017 (2009)
   [arXiv:0906.2435 [hep-ph]].

%\cite{Alexander:2016aln}
\bibitem{Alexander:2016aln} 
  J.~Alexander {\it et al.},
  %``Dark Sectors 2016 Workshop: Community Report,''
  arXiv:1608.08632 [hep-ph].

%\cite{Gninenko:2001hx}
\bibitem{Gninenko:2001hx} 
  S.~N.~Gninenko and N.~V.~Krasnikov,
  %``The Muon anomalous magnetic moment and a new light gauge boson,''
  Phys.\ Lett.\ B {\bf 513}, 119 (2001)
   [hep-ph/0102222].

%\cite{Baek:2001kca}
\bibitem{Baek:2001kca} 
  S.~Baek, N.~G.~Deshpande, X.~G.~He and P.~Ko,
  %``Muon anomalous g-2 and gauged L(muon) - L(tau) models,''
  Phys.\ Rev.\ D {\bf 64}, 055006 (2001)
  [hep-ph/0104141].

%\cite{Ma:2001md}
\bibitem{Ma:2001md} 
  E.~Ma, D.~P.~Roy and S.~Roy,
  %``Gauged L(mu) - L(tau) with large muon anomalous magnetic moment and the bimaximal mixing of neutrinos,''
  Phys.\ Lett.\ B {\bf 525}, 101 (2002)
  [hep-ph/0110146].

%\cite{Pospelov:2008zw}
\bibitem{Pospelov:2008zw} 
  M.~Pospelov,
  %``Secluded U(1) below the weak scale,''
  Phys.\ Rev.\ D {\bf 80}, 095002 (2009)
  [arXiv:0811.1030 [hep-ph]].

%\cite{Heeck:2011wj}
\bibitem{Heeck:2011wj} 
  J.~Heeck and W.~Rodejohann,
  %``Gauged L_mu - L_tau Symmetry at the Electroweak Scale,''
  Phys.\ Rev.\ D {\bf 84}, 075007 (2011)
  [arXiv:1107.5238 [hep-ph]].

%\cite{Krasnikov:2017dmg}
\bibitem{Krasnikov:2017dmg} 
  N.~V.~Krasnikov,
  %``The muon (g - 2) anomaly and a new light vector boson,''
  arXiv:1702.04596 [hep-ph].

%\cite{Fayet:2007ua}
\bibitem{Fayet:2007ua} 
  P.~Fayet,
  %``U-boson production in e+ e- annihilations, psi and Upsilon decays, and Light Dark Matter,''
  Phys.\ Rev.\ D {\bf 75}, 115017 (2007)
    [hep-ph/0702176 [hep-ph]].

%\cite{ArkaniHamed:2008qn}
\bibitem{ArkaniHamed:2008qn} 
  N.~Arkani-Hamed, D.~P.~Finkbeiner, T.~R.~Slatyer and N.~Weiner,
  %``A Theory of Dark Matter,''
  Phys.\ Rev.\ D {\bf 79}, 015014 (2009)
  [arXiv:0810.0713 [hep-ph]].

%\cite{Petraki:2014uza}
\bibitem{Petraki:2014uza} 
  K.~Petraki, L.~Pearce and A.~Kusenko,
  %``Self-interacting asymmetric dark matter coupled to a light massive dark photon,''
  JCAP {\bf 1407}, 039 (2014)
  [arXiv:1403.1077 [hep-ph]].

%%%%%new ref 23/03

%\cite{Foot:2014uba}
\bibitem{Foot:2014uba} 
  R.~Foot and S.~Vagnozzi,
  %``Dissipative hidden sector dark matter,''
  Phys.\ Rev.\ D {\bf 91}, 023512 (2015)
  [arXiv:1409.7174 [hep-ph]].


%%%%%%%%endnewrewf23/03

%\cite{Araki:2015mya}
\bibitem{Araki:2015mya} 
  T.~Araki, F.~Kaneko, T.~Ota, J.~Sato and T.~Shimomura,
  %``MeV scale leptonic force for cosmic neutrino spectrum and muon anomalous magnetic moment,''
  Phys.\ Rev.\ D {\bf 93},  013014 (2016)
   [arXiv:1508.07471 [hep-ph]].

%\cite{Baek:2015fea}
\bibitem{Baek:2015fea} 
  S.~Baek,
  %``Dark matter and muon $(g-2)$ in local $U(1)_{L_\mu-L_\tau}$-extended Ma Model,''
  Phys.\ Lett.\ B {\bf 756}, 1 (2016)
  [arXiv:1510.02168 [hep-ph]].

%\cite{Ko:2016uft}
\bibitem{Ko:2016uft} 
  P.~Ko and Y.~Tang,
  %``Light dark photon and fermionic dark radiation for the Hubble constant and the structure formation,''
  Phys.\ Lett.\ B {\bf 762}, 462 (2016)
  [arXiv:1608.01083 [hep-ph]].


%\cite{Chen:2016kxw}
\bibitem{Chen:2016kxw} 
  C.~H.~Chen and T.~Nomura,
  %``Light gauge boson in rare $K$ decay,''
  Phys.\ Lett.\ B {\bf 763}, 304 (2016)
  [arXiv:1608.02311 [hep-ph]].

%%%%%%%%new 23/03

%\cite{Campos:2017dgc}
\bibitem{Campos:2017dgc} 
  M.~D.~Campos, D.~Cogollo, M.~Lindner, T.~Melo, F.~S.~Queiroz and W.~Rodejohann,
  %``Neutrino Masses and Absence of Flavor Changing Interactions in the 2HDM from Gauge Principles,''
  JHEP {\bf 1708}, 092 (2017)
  [arXiv:1705.05388 [hep-ph]].

%%%%%%%%%end new 23/03


%\cite{Davoudiasl:2014kua}
\bibitem{Davoudiasl:2014kua} 
  H.~Davoudiasl, H.~S.~Lee and W.~J.~Marciano,
  %``Muon $g−2$, rare kaon decays, and parity violation from dark bosons,''
  Phys.\ Rev.\ D {\bf 89}, 095006 (2014)
  [arXiv:1402.3620 [hep-ph]].

%\cite{Aad:2015sva}
\bibitem{Aad:2015sva} 
  G.~Aad {\it et al.} [ATLAS Collaboration],
  %``Search for new light gauge bosons in Higgs boson decays to four-lepton final states in $pp$ collisions at $\sqrt{s}=8$ TeV with the ATLAS detector at the LHC,''
  Phys.\ Rev.\ D {\bf 92},  092001 (2015)
  [arXiv:1505.07645 [hep-ex]].

%\cite{Davoudiasl:2012ag}
\bibitem{Davoudiasl:2012ag} 
  H.~Davoudiasl, H.~S.~Lee and W.~J.~Marciano,
  %``'Dark' Z implications for Parity Violation, Rare Meson Decays, and Higgs Physics,''
  Phys.\ Rev.\ D {\bf 85}, 115019 (2012)
  [arXiv:1203.2947 [hep-ph]].

%\cite{Davoudiasl:2013aya}
\bibitem{Davoudiasl:2013aya} 
  H.~Davoudiasl, H.~S.~Lee, I.~Lewis and W.~J.~Marciano,
  %``Higgs Decays as a Window into the Dark Sector,''
  Phys.\ Rev.\ D {\bf 88},  015022 (2013)
    [arXiv:1304.4935 [hep-ph]].

%\cite{Lee:2013fda}
\bibitem{Lee:2013fda} 
  H.~S.~Lee and M.~Sher,
  %``Dark Two Higgs Doublet Model,''
  Phys.\ Rev.\ D {\bf 87},  115009 (2013)
  [arXiv:1303.6653 [hep-ph]].

%\cite{Curtin:2014cca}
\bibitem{Curtin:2014cca} 
  D.~Curtin, R.~Essig, S.~Gori and J.~Shelton,
  %``Illuminating Dark Photons with High-Energy Colliders,''
  JHEP {\bf 1502}, 157 (2015)
   [arXiv:1412.0018 [hep-ph]].

%\cite{Lee:2016ief}
\bibitem{Lee:2016ief} 
  H.~S.~Lee and S.~Yun,
  %``Mini force: The $(B-L)+xY$ gauge interaction with a light mediator,''
  Phys.\ Rev.\ D {\bf 93}, 115028 (2016)
   [arXiv:1604.01213 [hep-ph]].



%\cite{Krasznahorkay:2015iga}
\bibitem{Krasznahorkay:2015iga} 
  A.~J.~Krasznahorkay {\it et al.},
  %``Observation of Anomalous Internal Pair Creation in Be8 : A Possible Indication of a Light, Neutral Boson,''
  Phys.\ Rev.\ Lett.\  {\bf 116}, 042501 (2016)
   [arXiv:1504.01527 [nucl-ex]].



%\cite{Feng:2016jff}
\bibitem{Feng:2016jff} 
  J.~L.~Feng, B.~Fornal, I.~Galon, S.~Gardner, J.~Smolinsky, T.~M.~P.~Tait and P.~Tanedo,
  %``Protophobic Fifth-Force Interpretation of the Observed Anomaly in $^8$Be Nuclear Transitions,''
  Phys.\ Rev.\ Lett.\  {\bf 117}, 071803 (2016)
  [arXiv:1604.07411 [hep-ph]].

%\cite{Feng:2016ysn}
\bibitem{Feng:2016ysn} 
  J.~L.~Feng, B.~Fornal, I.~Galon, S.~Gardner, J.~Smolinsky, T.~M.~P.~Tait and P.~Tanedo,
  %``Particle physics models for the 17 MeV anomaly in beryllium nuclear decays,''
  Phys.\ Rev.\ D {\bf 95}, 035017 (2017)
  [arXiv:1608.03591 [hep-ph]].

%\cite{Krasznahorkay:2017qfd}
\bibitem{Krasznahorkay:2017qfd} 
  A.~J.~Krasznahorkay {\it et al.},
  %``New experimental results for the 17 MeV particle created in $^8Be$,''
  EPJ Web Conf.\  {\bf 137}, 08010 (2017).


%\cite{Krasznahorkay:2017gwn}
\bibitem{Krasznahorkay:2017gwn} 
  A.~J.~Krasznahorkay {\it et al.},
  %``On the creation of the 17 MeV X boson in the 17.6 MeV M1 transition of $^8Be$,''
  EPJ Web Conf.\  {\bf 142}, 01019 (2017).

%\cite{Gu:2016ege}
\bibitem{Gu:2016ege} 
  P.~H.~Gu and X.~G.~He,
  %``Realistic model for a fifth force explaining anomaly in ${^8Be^*} \to {^8Be} \;{e^+e^-}$ Decay,''
  Nucl.\ Phys.\ B {\bf 919}, 209 (2017)
  [arXiv:1606.05171 [hep-ph]].

%%%%new 1/02/2018 Boehm:2003hm,Borodatchenkova:2005ct

%\cite{Boehm:2003hm}
\bibitem{Boehm:2003hm} 
  C.~Boehm and P.~Fayet,
  %``Scalar dark matter candidates,''
  Nucl.\ Phys.\ B {\bf 683}, 219 (2004)
  [hep-ph/0305261].


%\cite{Borodatchenkova:2005ct}
\bibitem{Borodatchenkova:2005ct} 
  N.~Borodatchenkova, D.~Choudhury and M.~Drees,
  %``Probing MeV dark matter at low-energy e+e- colliders,''
  Phys.\ Rev.\ Lett.\  {\bf 96}, 141802 (2006)
  [hep-ph/0510147].

%\cite{Chen:2016dhm}
\bibitem{Chen:2016dhm} 
  L.~B.~Chen, Y.~Liang and C.~F.~Qiao,
  %``X(16.7) Production in Electron-Positron Collision,''
  arXiv:1607.03970 [hep-ph].
	
	

%%%%%%%end new

%\cite{Liang:2016ffe}
\bibitem{Liang:2016ffe} 
  Y.~Liang, L.~B.~Chen and C.~F.~Qiao,
  %``X(16.7) as the solution of the NuTeV anomaly,''
  Chin.\ Phys.\ C {\bf 41},  063105 (2017)
  [arXiv:1607.08309 [hep-ph]].

%\cite{Jia:2016uxs}
\bibitem{Jia:2016uxs} 
  L.~B.~Jia and X.~Q.~Li,
  %``The new interaction suggested by the anomalous $^8$Be transition sets a rigorous constraint on the mass range of dark matter,''
  Eur.\ Phys.\ J.\ C {\bf 76}, 706 (2016)
   [arXiv:1608.05443 [hep-ph]].

%\cite{Kitahara:2016zyb}
\bibitem{Kitahara:2016zyb} 
  T.~Kitahara and Y.~Yamamoto,
  %``Protophobic Light Vector Boson as a Mediator to the Dark Sector,''
  Phys.\ Rev.\ D {\bf 95}, 015008 (2017)
    [arXiv:1609.01605 [hep-ph]].

%\cite{Ellwanger:2016wfe}
\bibitem{Ellwanger:2016wfe} 
  U.~Ellwanger and S.~Moretti,
  %``Possible Explanation of the Electron Positron Anomaly at 17 MeV in $^8Be$ Transitions Through a Light Pseudoscalar,''
  JHEP {\bf 1611}, 039 (2016)
   [arXiv:1609.01669 [hep-ph]].

%\cite{Chen:2016tdz}
\bibitem{Chen:2016tdz} 
  C.~S.~Chen, G.~L.~Lin, Y.~H.~Lin and F.~Xu,
  %``The 17 MeV Anomaly in Beryllium Decays and $U(1)$ Portal to Dark Matter,''
  arXiv:1609.07198 [hep-ph].

%\cite{Seto:2016pks}
\bibitem{Seto:2016pks} 
  O.~Seto and T.~Shimomura,
  %``Atomki anomaly and dark matter in a radiative seesaw model with gauged $B-L$ symmetry,''
  Phys.\ Rev.\ D {\bf 95}, 095032 (2017)
   [arXiv:1610.08112 [hep-ph]].

%\cite{Kozaczuk:2016nma}
\bibitem{Kozaczuk:2016nma} 
  J.~Kozaczuk, D.~E.~Morrissey and S.~R.~Stroberg,
  %``Light axial vector bosons, nuclear transitions, and the $^8$Be anomaly,''
  Phys.\ Rev.\ D {\bf 95}, 115024 (2017)
  [arXiv:1612.01525 [hep-ph]].

%\cite{Zhang:2017zap}
\bibitem{Zhang:2017zap} 
  X.~Zhang and G.~A.~Miller,
  %``Can nuclear physics explain the anomaly observed in the internal pair production in the Beryllium-8 nucleus?,''
  arXiv:1703.04588 [nucl-th].

%\cite{DelleRose:2017xil}
\bibitem{DelleRose:2017xil} 
  L.~Delle Rose, S.~Khalil and S.~Moretti,
  %``Explanation of the 17 MeV Atomki Anomaly in a $U(1)^\prime$-Extended 2-Higgs Doublet Model,''
  arXiv:1704.03436 [hep-ph].

%\cite{Fornal:2017msy}
\bibitem{Fornal:2017msy} 
  B.~Fornal,
  %``Is There a Sign of New Physics in Beryllium Transitions?,''
  arXiv:1707.09749 [hep-ph].

%%%%%% new cite 23/03

%\cite{Banerjee:2018vgk}
\bibitem{Banerjee:2018vgk} 
  D.~Banerjee {\it et al.} [NA64 Collaboration],
  %``Search for a new X(16.7) boson and dark photons in the NA64 experiment at CERN,''
  arXiv:1803.07748 [hep-ex].

%%%%% end new cite 23/03




%\cite{Batley:2015lha}
\bibitem{Batley:2015lha} 
  J.~R.~Batley {\it et al.} [NA48/2 Collaboration],
  %``Search for the dark photon in $\pi^0$ decays,''
  Phys.\ Lett.\ B {\bf 746}, 178 (2015)
   [arXiv:1504.00607 [hep-ex]].



%%%%%%%%%%%%%%%%new cite 13/03

%\cite{Essig:2009nc}
\bibitem{Essig:2009nc} 
  R.~Essig, P.~Schuster and N.~Toro,
  %``Probing Dark Forces and Light Hidden Sectors at Low-Energy e+e- Colliders,''
  Phys.\ Rev.\ D {\bf 80}, 015003 (2009)
  [arXiv:0903.3941 [hep-ph]].

%\cite{Essig:2013vha}
\bibitem{Essig:2013vha} 
  R.~Essig, J.~Mardon, M.~Papucci, T.~Volansky and Y.~M.~Zhong,
  %``Constraining Light Dark Matter with Low-Energy $e^+e^-$ Colliders,''
  JHEP {\bf 1311}, 167 (2013)
  [arXiv:1309.5084 [hep-ph]].

%\cite{Lees:2014xha}
\bibitem{Lees:2014xha} 
  J.~P.~Lees {\it et al.} [BaBar Collaboration],
  %``Search for a Dark Photon in $e^+e^-$ Collisions at BaBar,''
  Phys.\ Rev.\ Lett.\  {\bf 113}, 201801 (2014)
  [arXiv:1406.2980 [hep-ex]].

%\cite{Lees:2017lec}
\bibitem{Lees:2017lec} 
  J.~P.~Lees {\it et al.} [BaBar Collaboration],
  %``Search for Invisible Decays of a Dark Photon Produced in ${e}^{+}{e}^{-}$ Collisions at BaBar,''
  Phys.\ Rev.\ Lett.\  {\bf 119}, 131804 (2017)
  [arXiv:1702.03327 [hep-ex]].


%%%%%%%%%%%%%%%%end new cite 13/03



%\cite{Patrignani:2016xqp}
\bibitem{Patrignani:2016xqp} 
  C.~Patrignani {\it et al.} [Particle Data Group],
  %``Review of Particle Physics,''
  Chin.\ Phys.\ C {\bf 40},  100001 (2016).



%\cite{Deniz:2009mu}
\bibitem{Deniz:2009mu} 
  M.~Deniz {\it et al.} [TEXONO Collaboration],
  %``Measurement of Nu(e)-bar -Electron Scattering Cross-Section with a CsI(Tl) Scintillating Crystal Array at the Kuo-Sheng Nuclear Power Reactor,''
  Phys.\ Rev.\ D {\bf 81}, 072001 (2010)
  [arXiv:0911.1597 [hep-ex]].

%%%%%%%%%%%%%%%new reference

%\cite{Nardi:2018cxi}
\bibitem{Nardi:2018cxi} 
  E.~Nardi, C.~D.~R.~Carvajal, A.~Ghoshal, D.~Meloni and M.~Raggi,
  %``Resonant production of dark photons in positron beam dump experiments,''
  arXiv:1802.04756 [hep-ph].

%\cite{Araki:2017wyg}
\bibitem{Araki:2017wyg} 
  T.~Araki, S.~Hoshino, T.~Ota, J.~Sato and T.~Shimomura,
  %``Detecting the $L_{\mu}-L_{\tau}$ gauge boson at Belle II,''
  Phys.\ Rev.\ D {\bf 95}, 055006 (2017)
  [arXiv:1702.01497 [hep-ph]].
	
	

%\cite{Chen:2017cic}
\bibitem{Chen:2017cic} 
  C.~H.~Chen and T.~Nomura,
  %``$L_\mu -L_\tau$ gauge-boson production from lepton flavor violating $\tau$ decays at Belle II,''
  Phys.\ Rev.\ D {\bf 96}, 095023 (2017)
    [arXiv:1704.04407 [hep-ph]].

%\cite{Gninenko:2017yus}
\bibitem{Gninenko:2017yus} 
  S.~N.~Gninenko, D.~V.~Kirpichnikov, M.~M.~Kirsanov and N.~V.~Krasnikov,
  %``The exact tree-level calculation of the dark photon production in high-energy electron scattering at the CERN SPS,''
  arXiv:1712.05706 [hep-ph].



\bibitem{byckling} 
See, for example,~(2.7) of Section II in E.~Byckling and K.~Kajantie, Particle Kinematics, Wiley, London, 1973.



%\cite{Aubert:2001tu}
\bibitem{Aubert:2001tu}
  B.~Aubert {\it et al.} [BaBar Collaboration],
  %``The BaBar detector,''
  Nucl.\ Instrum.\ Meth.\ A {\bf 479}, 1 (2002)
  [hep-ex/0105044].
	
	
	
%\cite{Bjorken:2009mm}
\bibitem{Bjorken:2009mm} 
  J.~D.~Bjorken, R.~Essig, P.~Schuster and N.~Toro,
  %``New Fixed-Target Experiments to Search for Dark Gauge Forces,''
  Phys.\ Rev.\ D {\bf 80}, 075018 (2009)
   [arXiv:0906.0580 [hep-ph]].


%\cite{Bossi:2013lxa}
\bibitem{Bossi:2013lxa} 
  F.~Bossi,
  %``Dark Photon Searches Using Displaced Vertices at Low Energy $e^+ e^-$ Colliders,''
  Adv.\ High Energy Phys.\  {\bf 2014}, 891820 (2014)
  [arXiv:1310.8181 [hep-ex]].

%\cite{Chen:1975sh}
\bibitem{Chen:1975sh} 
  M.~S.~Chen and P.~M.~Zerwas,
  %``Equivalent-Particle Approximations in electron and Photon Processes of Higher Order QED,''
  Phys.\ Rev.\ D {\bf 12}, 187 (1975).
	
	%\cite{Riordan:1987aw}
\bibitem{Riordan:1987aw} 
  E.~M.~Riordan {\it et al.},
  %``A Search for Short Lived Axions in an Electron Beam Dump Experiment,''
  Phys.\ Rev.\ Lett.\  {\bf 59}, 755 (1987).



\end{thebibliography}
\end{document}